# Beyond the Local Volume. I. Surface Densities of Ultracool Dwarfs in Deep HST/WFC3 Parallel Fields

Christian Aganze[1], Adam J. Burgasser[1], Mathew Malkan[2], Christopher A. Theissen[1,7], Roberto A. Tejada Arevalo[3], Chih-Chun Hsu[1], Daniella C. Bardalez Gagliuffi[4], Russell E. Ryan, Jr.[5], and Benne Holwerda[6]
[1] Department of Physics & Astronomy, University of California, San Diego, CA 92093, USA; caganze@ucsd.edu
[2] Department of Physics & Astronomy, University of California, Los Angeles, CA 90095, USA
[3] Department of Astrophysical Sciences, Princeton University, Princeton, NJ 08544, USA
[4] Department of Astrophysics, American Museum of Natural History, New York, NY 10024, USA
[5] Space Telescope Science Institute, Baltimore, MD 21218, USA
[6] Department of Physics and Astronomy, University of Louisville, Louisville, KY 40292, USA

## Abstract

Ultracool dwarf stars and brown dwarfs provide a unique probe of large-scale Galactic structure and evolution; however, until recently spectroscopic samples of sufficient size, depth, and fidelity have been unavailable. Here, we present the identification of 164 M7-T9 ultracool dwarfs in 0.6 $deg^2$ of deep, low-resolution, near-infrared spectroscopic data obtained with the Hubble Space Telescope (HST) Wide Field Camera 3 (WFC3) instrument as part of the WFC3 Infrared Spectroscopic Parallel Survey and the 3D-HST survey. We describe the methodology by which we isolate ultracool dwarf candidates from over 200,000 spectra, and show that selection by machine-learning classification is superior to spectral index-based methods in terms of completeness and contamination. We use the spectra to accurately determine classifications and spectrophotometric distances, the latter reaching to ∼2 kpc for L dwarfs and ∼400 pc for T dwarfs.

## 1. Introduction

The structure and evolution of the Milky Way is largely informed from the spatial and kinematic distributions of its luminous stars, an area of study known as Galactic archeology (Freeman 1987; Ivezić et al. 2012). Large imaging, astrometric, and spectroscopic surveys such as the Sloan Digital Sky Survey (SDSS; York et al. 2000) have been critical for building our model of the Milky Way through six-dimensional position and velocity data, and detailed spectroscopic characterization of large-scale stellar populations. Star-count data spanning decades have revealed that the Milky Way contains multiple stellar populations, including a kinematically young population spatially distributed into two exponential disks (the thin and thick disks); a centrally concentrated and spherically distributed population of metal-rich stars (the bulge), and a widely dispersed, old, and metal-depleted population (the halo; de Vaucouleurs & Pence 1978; Bahcall & Soneira 1981; Jurić et al. 2008). Analysis of the stellar evolutionary states of these populations suggest that the Galactic disk population has been continuously forming stars over the past 8–11 Gyr, while halo stars were largely formed 10–13 Gyr ago (Leggett et al. 1998; Tolstoy et al. 2009; Haywood et al. 2013).

More recently, the Gaia astrometric mission (Gaia Collaboration et al. 2018), combined with large-scale spectroscopic surveys such as SDSS, the Large sky Area Multi-Object fiber Spectroscopic Telescope (Zhao et al. 2012), the Apache Point Observatory Galactic Evolution Experiment (Majewski et al. 2017), and Galactic Archaeology with HERMES (De Silva et al. 2015; Martell et al. 2017), has greatly refined this picture of the Milky Way. Notable insights include the inference of major merger events that likely formed the inner stellar halo and thick disk populations (Gaia-Enceladus/Gaia sausage: Belokurov et al. 2018; Helmi et al. 2018; Myeong et al. 2018; Gallart et al. 2019; and the Sequoia event: Myeong et al. 2018, 2019); phase-space substructure and mixing among stars in the solar neighborhood, indicative of past perturbations of the Milky Way by satellite systems (Antoja et al. 2018); an ensemble of stellar streams that trace the tidal disruption of, and stellar accretion from, these satellites (Boubert et al. 2018; Malhan et al. 2018; Koppelman et al. 2019); and the detection of dozens of hypervelocity stars ejected through encounters with central supermassive black holes in the Milky Way and Large Magellanic Cloud (Boubert et al. 2018; Erkal et al. 2019). These results show the Milky Way and its environment to be a complex and dynamically evolving system.

All of these studies have focused on the brightest red giants and main-sequence stars, which provide both reach and reliability in the inference of stellar properties. Ultracool dwarfs (UCDs; $M \lesssim 0.1\ M_{\odot}$, $T_{eff} \lesssim 3000$ K; Kirkpatrick 2005) provide an alternative, and potentially more enriching, population for studying the Milky Way system (Burgasser 2004; Ryan et al. 2017). UCDs constitute ∼50% of stars by number in the immediate solar neighborhood ($d < 100$ pc), and are abundant in every environment in the Galaxy (Reid et al. 1999; Chabrier & Baraffe 2000; Cruz et al. 2007; Bochanski et al. 2010; Kirkpatrick et al. 2019). Stellar UCDs have lifetimes far in excess of the age of the Galaxy ($>10^3$ Gyr, Laughlin et al. 1997), while substellar UCDs (brown dwarfs) do not fuse hydrogen and have effectively limitless lifetimes (Kumar 1962, 1963; Hayashi & Nakano 1963). Brown dwarfs also cool and dim as they age, developing distinct spectra shaped by strong molecular absorption features that are highly sensitive to atmospheric temperature, surface gravity, and metallicity. The thermal and chemical evolution of stellar and substellar UCDs provide age diagnostics that have been exploited in stellar cluster studies (Stauffer et al. 1998; Jeffries & Oliveira 2005; Martín et al. 2018), coeval binary systems (Song et al. 2002; Burgasser & Blake 2009), and searches of new members of young moving groups near the Sun (Lopez-Santiago et al. 2006; Gagné et al. 2015; Mamajek et al. 2015; Faherty et al. 2018).

[7] NASA Sagan Fellow.

UCDs have historically been uncovered in wide-field red optical and infrared sky surveys, including SDSS, the Deep Near-Infrared Survey of the Southern Sky (Epchtein et al. 1997), the Two Micron All-Sky Survey (2MASS; Skrutskie et al. 2006), the United Kingdom Infrared Telescope Infrared Deep Sky Survey (UKIDSS; Lawrence et al. 2007), the Canada-France Brown Dwarf Survey (Reylé et al. 2010), the Wide-field Infrared Survey Explorer (WISE; Wright et al. 2010), and the Panoramic Survey Telescope and Rapid Response System (Chambers et al. 2016), among others. The intrinsic faintness of UCDs means that these surveys only reach the immediate solar neighborhood ($d \leqslant 100$ pc). While these surveys have enabled study of the local UCD luminosity and mass functions (Cruz et al. 2007; Metchev et al. 2008; Reyle et al. 2010; Bardalez Gagliuffi et al. 2019; Kirkpatrick et al. 2019, 2021) they cannot probe Galactic structure or UCD halo populations, of which relatively few examples are currently known (Burgasser et al. 2003; Lépine & Scholz 2008; Zhang et al. 2019).

Deep, narrow-field imaging surveys provide one approach to investigating more distant UCD populations. Many of these surveys have exploited the sensitivity and imaging resolution of the Hubble Space Telescope (HST), often in parallel with searches for high-redshift galaxies for which UCDs are a *contaminant* population (Reid et al. 1996; Ryan et al. 2005, 2011; Holwerda et al. 2014). Very deep, multiband, ground-based surveys have also expanded our reach for UCDs (Kakazu et al. 2010; Carnero Rosell et al. 2019; Sorahana et al. 2019). However, these imaging surveys are subject to contamination and inaccuracies in spectral classification, which inhibits a detailed evaluation of completeness and the analysis of population composition in terms of mass, temperature, metallicity, and other properties. Deep spectral surveys, notably those deploying the slitless grism modes of the HST/Advanced Camera for Surveys (ACS) and HST/Wide Field Camera 3 (WFC3) instruments, have provided more robust and well-characterized samples of distant UCDs, but are shallower and smaller in sample size ($\lesssim$50 sources; Pirzkal et al. 2005, 2009; Masters et al. 2012).

In this paper, we present a new analysis of HST/WFC3 slitless grism spectra from the WFC3 Infrared Spectroscopic Parallel Survey (WISPS; Atek et al. 2010) and 3D-HST (Momcheva et al. 2016; Brammer et al. 2012; Skelton et al. 2014) survey, expanding both the areal coverage and spectral types evaluated in these deep spectroscopic data sets. In Section 2, we describe the survey data used. In Section 3, we describe our selection process, including a robust analysis of three independent procedures using spectral indices and template fitting: selection by predefined index ranges and by two supervised machine-learning methods. In Section 4, we review the properties of the 164 UCDs identified in this sample, including classifications and spectrophotometric distance estimates. In Section 5, we summarize the main results of our study. Further analysis of this sample is presented in a companion paper (C. A. Aganze et al. 2021, in preparation; hereafter Paper II).

## 2. Data

The WISPS and 3D-HST surveys used the infrared channel of the WFC3 camera (Kimble et al. 2008), obtaining low-resolution G102 ($\lambda = 0.8$–$1.17\,\mu$m, $\lambda/\Delta\lambda \approx 200$) and G141 ($\lambda = 1.11$–$1.67\,\mu$m, $\lambda/\Delta\lambda \approx 130$) grism spectra. Removal of the slit mask allows for overlapping spectra across the $136'' \times 123''$ ($3.1 \times 10^{-4}$ arcmin$^2$) inner field of view of the WFC3 camera.

### 2.1. WISPS Survey Data

WISPS is a 1000 orbit, HST pure-parallels survey covering 390 fields ($\sim$1500 arcmin$^2$), obtained concurrent with observations made with the Cosmic Origins Spectrograph (COS) or Space Telescope Imaging Spectrograph (STIS). The goal of WISPS is to conduct a census of star-forming, high-redshift galaxies; hence, WISPS fields are typically at high Galactic latitudes ($|b| \gtrsim 20°$). Grism data in G102 and G141 settings were obtained with an exposure time ratio of 2.4:1, although the individual exposure times varied. For G141 data, exposures ranged over 600–1400 s and 3000–9000 s for G102 data. Data were reduced using a combination of `aXe` (Kümmel et al. 2009) and custom software, the latter to remove residual background and to flag bad pixels. Reference images were used to determine source location, and contamination corrections for overlapping spectra were computed from `aXe`. Direct images were also obtained in the broadband F110W, F140W, and F160W filters, with source catalogs generated using SExtractor (Bertin & Arnouts 1996). We analyzed reduced grism and photometric data provided in WISPS release version 6.2 (Atek et al. 2010).

### 2.2. 3D-HST Survey Data

3D-HST is an HST parallels survey of 248 orbits covering $\sim$600 arcmin$^2$. The goal of 3D-HST is to understand the physical processes that shape galaxies in the high-redshift universe. This survey targeted four standard deep extragalactic fields: the All-wavelength Extended Groth strip International Survey (AEGIS; Davis et al. 2007), the Cosmic Evolution Survey (COSMOS; Scoville et al. 2007); the UKIDSS Ultra-Deep Survey (UKIDSS-UDS; Lawrence et al. 2007), and the Great Observatories Origins Deep Survey (GOODS-South and GOODS-North; Giavalisco et al. 2004). These fields are also covered by the Cosmic Assembly Near-infrared Deep Extragalactic Legacy Survey (Grogin et al. 2011; Koekemoer et al. 2011). Spectral data were obtained using both the ACS/G800L grism covering 0.5–0.9 $\mu$m with a resolution of $\lambda/\Delta\lambda \approx 100$, and the WFC3/G141 grism. We focused on the WFC3 data for this study. WFC3 exposures were obtained over two orbits, with exposure times varying between 2500 and 6600 s per pointing. Data were reduced using custom software, including generation of a contamination model, as described in Momcheva et al. (2016); we used these data for our spectral analysis.[8] Note that the extracted 1D spectra reported in

[8] Data were retrieved from the survey's website, https://3dhst.research.yale.edu/Home.html, on 2016 August 12.

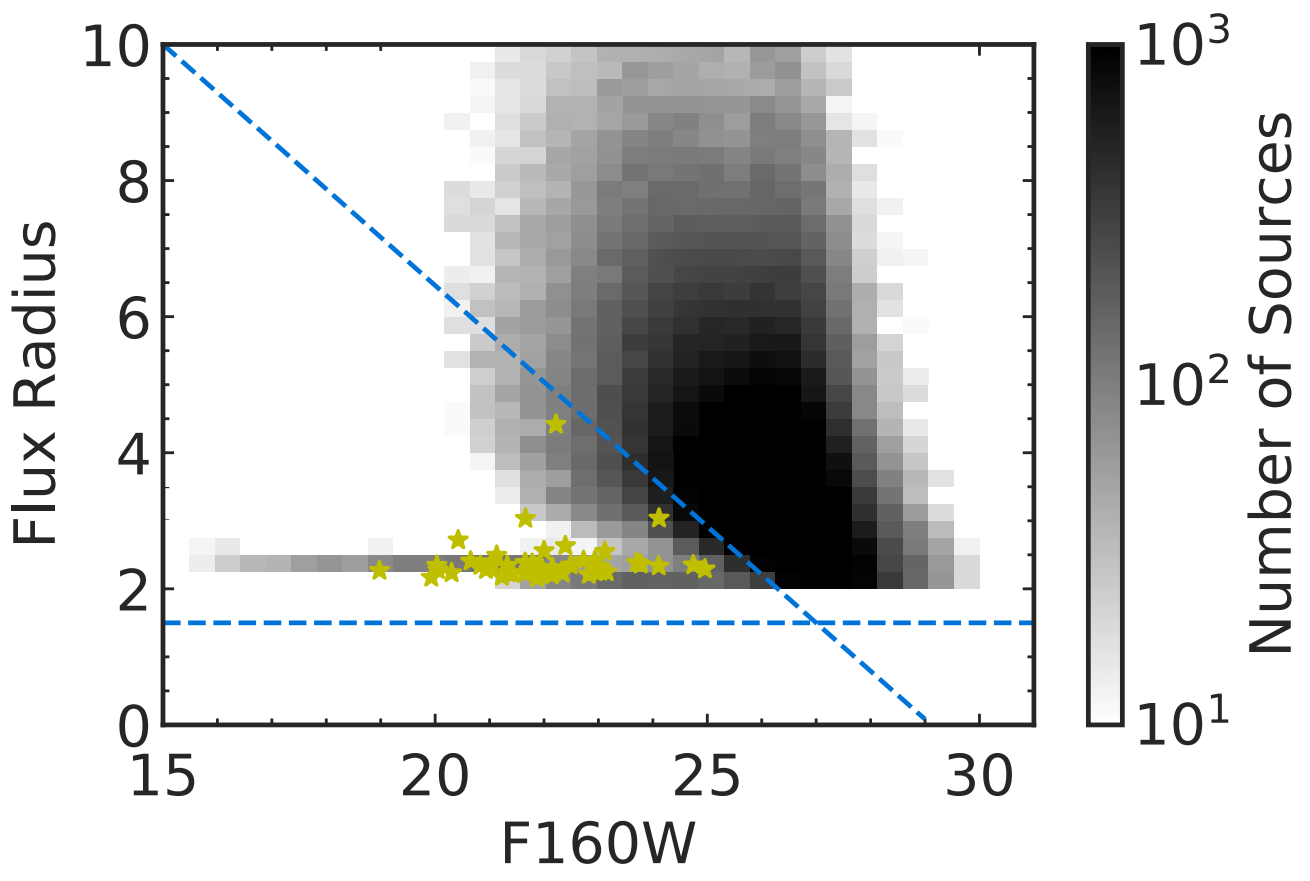

**Figure 1.** Point-source selection cut for the 3D-HST survey. The shaded 2D histogram shows the density of sources in 3D-HST, while our final sample is shown as yellow stars set by the criteria $r \geqslant 1.5$ and $r < -0.7 \times$ F140W + 20.6.

Momcheva et al. (2016) are not corrected for sensitivity; we applied the sensitivity curves provided in the same 3D-HST data products to recover each spectrum's continuum shape.

## 3. Selection of UCDs

### 3.1. Spectral Calibration Sample

To anchor our identification of new UCDs in the WISPS and 3D-HST data sets, we first created a spectral calibration sample composed of known UCDs with similar spectral coverage and resolution to the HST spectra. This sample was drawn from nearly 3000 low-resolution ($\lambda/\Delta\lambda \approx 75$–120), near-infrared (0.9–2.5 $\mu$m) spectra of nearby UCDs observed with the SpeX spectrograph on the NASA InfraRed Telescope Facility (Rayner et al. 2003), all contained in the SpeX Prism Library (Burgasser 2014). We selected all spectra with median signal-to-noise ratio (S/N) > 10, and visually inspected these spectra to remove background contaminants, which are retained for our random forest classification as non-UCDs (Section 3.4). We also used 22 Y dwarf spectra from Schneider et al. (2015) and 77 UCD spectra from Manjavacas et al. (2019), all obtained with HST/WFC3. The resulting sample of 2,197 spectra of UCDs with spectral types M7 and later is referred hereafter as the UCD spectral template sample.

### 3.2. Preselection Constraints

#### 3.2.1. Point-source Selection

We combined all grism and photometric data from the WISPS and 3D-HST data sets, an initial sample of 254,264 sources. To refine our selection of UCDs, point sources in the WISPS survey were selected using the SExtractor stellarity index `CLASS_STAR` ≠ 0. 3D-HST provides a different flag, `star_flag`, that identifies point sources based on F160W imaging data. However, this flag rejected three visually confirmed UCDs from the 3D-HST sample that were not rejected with the SExtractor stellarity index, so we used a different selection criteria based on the half-light radius parameter `FLUX_RADIUS` ($r$). Following Skelton et al. (2014), we required $r \geqslant 1.5$ and $r < -0.7 \times \mathrm{F140W} + 20.6$ (Figure 1). Point-source selection reduced the sample down to 104,346 sources, or 41% of the initial spectral sample.

#### 3.2.2. J-band S/N Rejection

To eliminate low S/N spectra that may be too ambiguous to identify or classify as UCDs, we applied an S/N cut on the WFC3 spectral data. UCDs display strong $H_2O$ and $CH_4$ molecular absorption features in the $J$ and $H$ bands (1.1–1.6 $\mu$m), so an S/N calculation that encompasses the full spectral range will produce different results for different spectral types. Therefore, we defined S/N ratios for the $J$-band continuum in the range 1.2 $\mu$m $\leqslant \lambda \leqslant$ 1.3 $\mu$m (hereafter $J$ S/N), the $H$-band continuum in the range 1.52 $\mu$m $\leqslant \lambda \leqslant$ 1.65 $\mu$m (hereafter $H$ S/N), and an overall median S/N in the range 1.1 $\mu$m $\leqslant \lambda \leqslant$ 1.65 $\mu$m (hereafter MEDIAN S/N). We also computed an average of the $J$-band and $H$-band S/N (hereafter $JH$ S/N). Figure 2 shows that a $J$ S/N cut at 3 allows us to effectively probe a deeper sample of later-type objects. We rejected the lowest S/N spectra by requiring $J$ S/N $\geqslant 3$, which retained 46,561 spectra, or 18% of the initial spectral sample.

#### 3.2.3. Spectral Template and Line Fitting

Visual inspection of the remaining spectral data identify some common contaminants, including emission line sources, featureless spectra with low S/N, spectra with absorption or emission features outside the primary $H_2O$ and $CH_4$ bands, and other artifacts. To distinguish featureless spectra from UCDs, we fit each WFC3 spectrum to the full UCD spectral template sample and to a straight line using $\chi^2$ minimization. The $\chi^2$ for each template fit was computed as

$$\chi_T^2 = \sum_{\lambda=1.15\,\mu\mathrm{m}}^{1.65\,\mu\mathrm{m}} \frac{[\mathrm{Sp}(\lambda) - \alpha T(\lambda)]^2}{\sigma(\lambda)^2}, \tag{1}$$

where Sp($\lambda$) is the WFC3 spectrum, $\sigma(\lambda)$ is its uncertainty, $T(\lambda)$ is the UCD template spectrum, and $\alpha$ is a scale factor that minimizes $\chi_T^2$,

$$\alpha = \frac{\sum_{\lambda=1.15\,\mu\mathrm{m}}^{1.65\,\mu\mathrm{m}} \frac{\mathrm{Sp}(\lambda) \times T(\lambda)}{\sigma(\lambda)^2}}{\sum_{\lambda=1.15\,\mu\mathrm{m}}^{1.65\,\mu\mathrm{m}} \frac{T(\lambda)^2}{\sigma(\lambda)^2}} \tag{2}$$

(see Cushing et al. 2005). The $\chi^2$ for each line fit was similarly computed as

$$\chi_L^2 = \sum_{\lambda=1.15\,\mu\mathrm{m}}^{1.65\,\mu\mathrm{m}} \frac{[\mathrm{Sp}(\lambda) - a - b\lambda]^2}{\sigma(\lambda)^2}, \tag{3}$$

where $a$ and $b$ are linear parameters determined through least-squares minimization.

To distinguish between these fits, we used an $F$-test statistic to determine if the best-fit template was a significant improvement over the line fit, by requiring $F(\chi_T^2/\chi_L^2, \mathrm{DOF}_T, \mathrm{DOF}_L) < 0.02$, where DOF is the effective degrees of freedom of each fit, equal to the number of pixels in the wavelength range 1.15–1.65 $\mu$m ($N = 108$), minus 1 for the template fit or minus 2 for the line fit. This constraint imposes the condition that the probability that the UCD template is a worse fit than a line is less than 0.5%, a limit chosen based on the distribution of $F$ statistics for similar fits to the spectral templates. This selection cut retained 98% of the templates and 5946 of the WISPS/3D-HST point sources (2.3% of the initial sample), and provided a first-order estimate of the spectral classification of true UCDs in our sample.

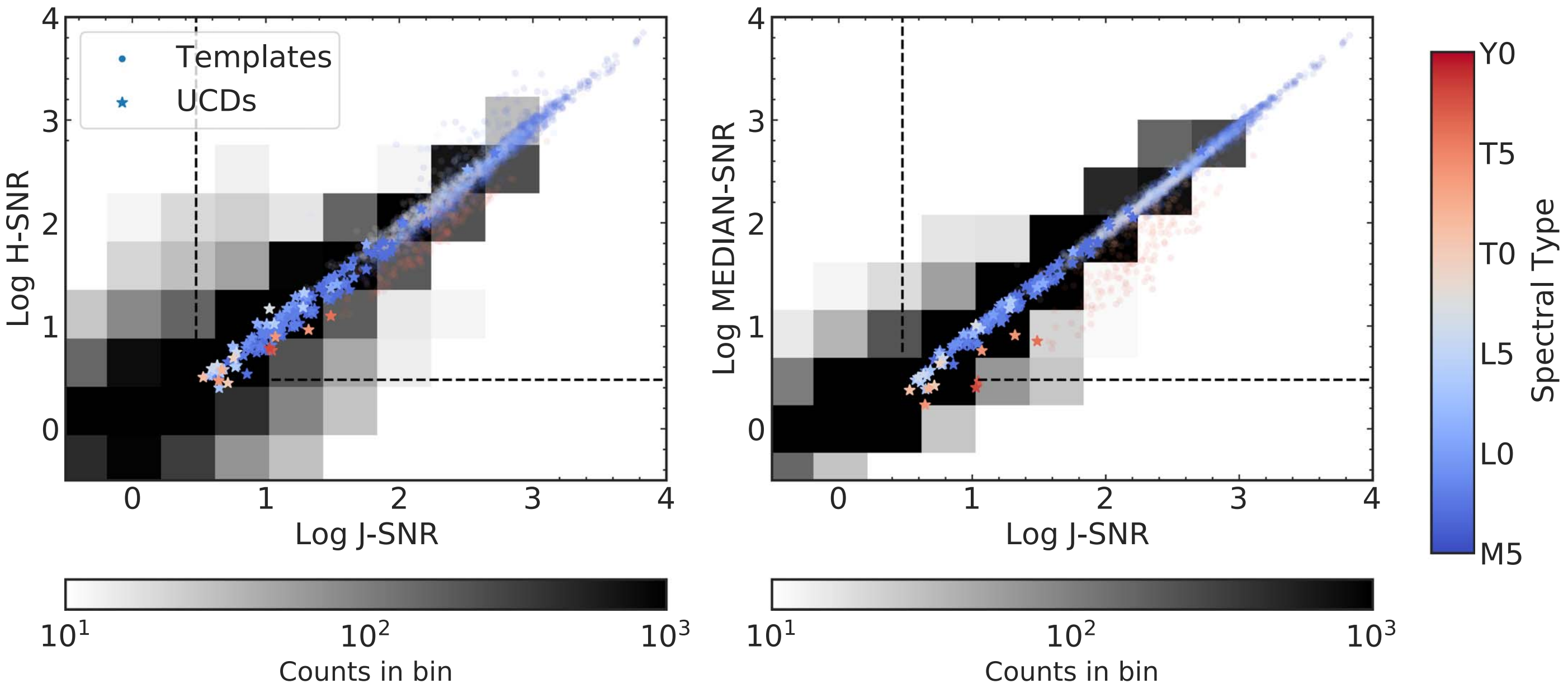

**Figure 2.** Comparison between $H$ S/N, MEDIAN S/N, and $J$ S/N for objects in our final sample of UCDs (stars), point sources in both surveys (gray background) and objects in our calibration sample colored by spectral type. Making a cut in $J$ S/N allows for a selection of late-type objects that would be missed from a comparable MEDIAN-S/N cut. Dashed lines indicate an S/N of 3.

**Table 1**
Spectral Indices

| Band | Wavelength Range ( $\mu$m) |
|---|---|
| J-cont | 1.15–1.20 |
| $H_2O$-1 | 1.246–1.295 |
| $H_2O$-2 | 1.38–1.43 |
| H-cont | 1.56–1.61 |
| $CH_4$ | 1.62–1.67 |

### 3.3. Spectral Index Selection

#### 3.3.1. Defining Spectral Indices

Our final selection criteria was based on the measurement of spectral indices which sample the strong $H_2O$ and $CH_4$ molecular features present in the $J$- and $H$-band regions. Spectral indices are commonly used to classify UCD spectra, and are typically tailored to the spectral resolution and region sampled (e.g., Tokunaga & Kobayashi 1999; Cushing et al. 2000; Testi et al. 2001; Allers et al. 2007; Burgasser et al. 2007). We selected five wavelength regions bracketing the primary absorption features (Table 1, Figure 3), measuring the median flux in these regions. Uncertainties were estimated by Monte Carlo sampling of the individual flux measurements, assuming normal distributions scaled by the spectral uncertainties. Ten index ratios were then defined from these band fluxes, each of the form

$$A/B = \frac{\langle F(\lambda_{A,l} < \lambda < \lambda_{A,h})\rangle}{\langle F(\lambda_{B,l} < \lambda < \lambda_{B,h})\rangle}, \quad (4)$$

where $\langle F\rangle$ denotes the median flux value in the wavelength range $\lambda_l < \lambda < \lambda_h$. To select for Y dwarfs, which have extremely low levels of flux in the $H_2O$ and $CH_4$ absorption bands, we defined six additional indices that averaged flux measurements in multiple absorption bands, e.g., ($H_2O$-1+$CH_4$)/$J$-cont, ($H_2O$-2+$CH_4$)/$H$-cont. Figure 4 displays the trends in these spectral indices among dwarf and subdwarf spectral standards.

The spectral indices were measured on both the WFC3 data and our UCD spectral template sample. The latter were used to determine selection criteria for UCD subtype groups of M7–M9, L0–L4, L5–L9, T0–T4, T5–T9, Y dwarfs, and late-M/L subdwarfs. For each possible index-index pairing, we evaluated the distribution of index measurements for calibration templates within each of these subtype groupings, and defined parallelogram regions in which these templates clustered. The parallelograms were determined by first measuring a linear trend ($I_2 = aI_1 + b$) in the index-index data for a given subtype group, then defining a perpendicular range about that line that encompasses five times the standard deviation of template values. This process produces four vertices encompassing each template cluster in each of the possible index-index spaces.

#### 3.3.2. Completeness and Contamination

To quantify the effectiveness of these selection regions to identify UCDs in a given subtype group and exclude contaminants, we defined two statistical metrics measuring completeness (CP) and contamination (CT):

$$\mathrm{CP(SG},\,R) = \frac{N_{\mathrm{T}}^{*}(\mathrm{SG},\,R)}{N_{\mathrm{T}}(\mathrm{SG})}, \quad (5)$$

$$\mathrm{CT}(R) = \frac{N_{\mathrm{WFC3}}^{*}(R)}{N_{\mathrm{WFC3}}}. \quad (6)$$

In Equation (5), $N_{\mathrm{T}}(\mathrm{SG})$ is the total number of templates in subtype group SG, while $N_{\mathrm{T}}^{*}(\mathrm{SG}, R)$ is the number of templates within an index-index selection region $R$; CP = 1 indicates selection of all templates. In Equation (6), $N_{\mathrm{WFC3}}$ is the total number of WFC3 spectra, while $N_{\mathrm{WFC3}}^{*}(R)$ is the number of these spectra within an index-index selection region $R$. As our expected number of UCD discoveries is assumed to be much smaller than $N_{\mathrm{WFC3}}$, a high selection rate of WFC3 spectra is consistent with a high level of contamination; CT $\ll$1 indicates minimum contamination.

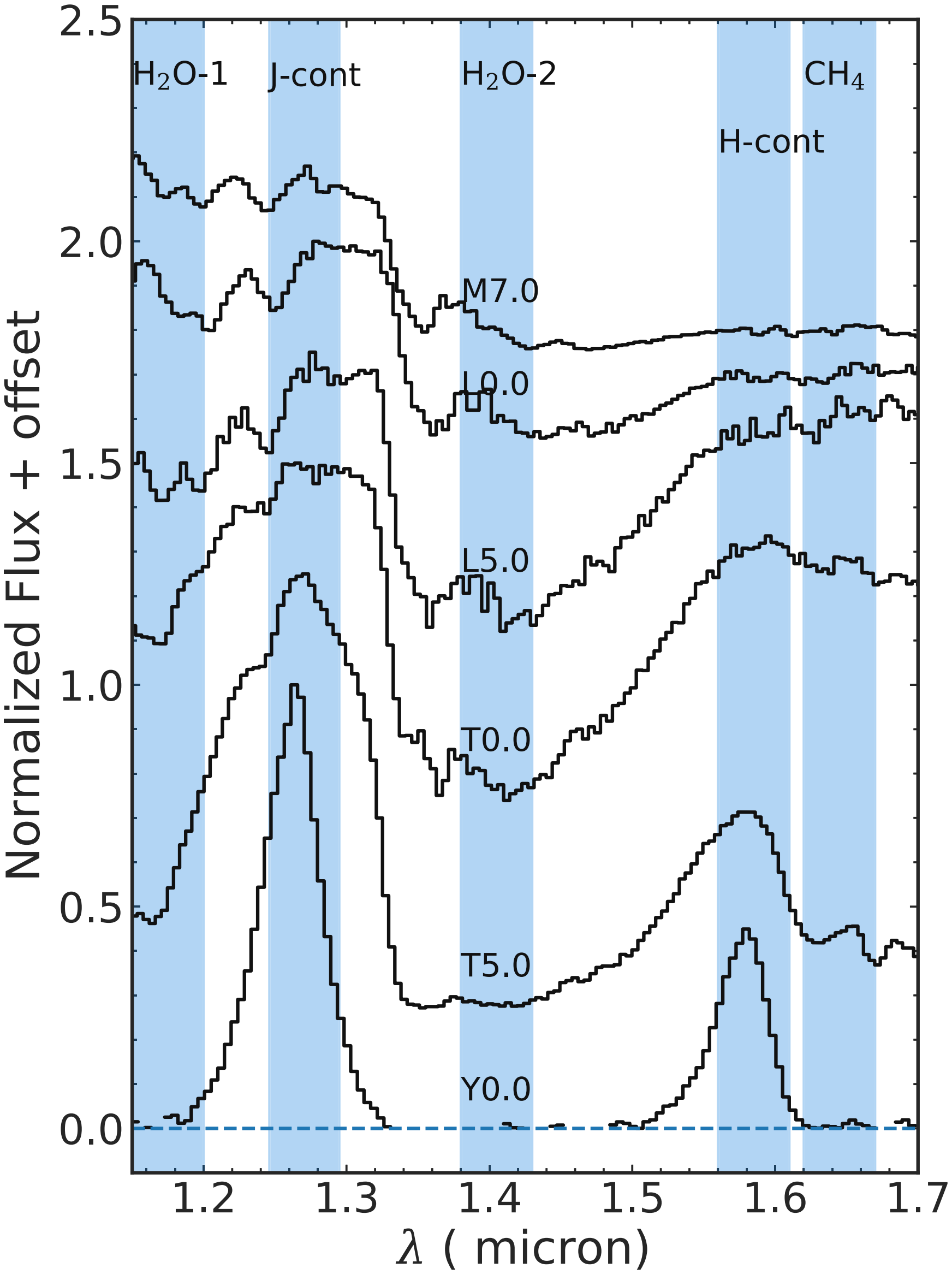

**Figure 3.** Illustration of spectral bands used to define spectral indices in this study (blue bands). The black lines are normalized low-resolution late-M, L, and T dwarf spectral standards defined in Burgasser et al. (2006a) and Kirkpatrick et al. (2010), and the Y0 dwarf WISE J1738+2732 from Cushing et al. (2011).

To maximize UCD selection (high CP) and minimize contamination (low CT) for each spectral subgroup, we rank ordered by CT those index-index pairs with CP $\geqslant$ 0.9, and chose the pair with the lowest contamination value; these are listed in Table 2. The best indices for each subtype group generally reflects the strongest molecular features present in those spectra (Figure 5). The greatest contamination was found in the selection regions for late-type M and early-type L dwarfs, and late-M/L subdwarfs, due to the relative weakness of molecular absorption features for these spectral types and hence greater similarity to background objects. Conversely, the late-type T and Y dwarf groups had the lowest contamination due to their distinct spectral morphologies.

Applying the single best index-index selection criteria to the sources that passed the prior cuts yielded 3400 unique objects, primarily identified as late-M dwarfs and subdwarfs (Table 2). The subdwarf selection criterion appear to be the most highly contaminated, as it identified 2042 objects, while visual inspection ruled out the detection of any unambiguous ultracool subdwarfs (Section 4.5).

### 3.3.3. Visual Selection

After the preceding selection criteria were applied, we visually inspected the remaining spectra to confirm their UCD nature and spectral classifications. Keeping only those sources whose spectra were a clear visual match to an UCD spectral standard (including d/sd and sd subdwarf standards), we identified a total of 164 UCDs, listed in Tables 3 and 4.

Retrospectively, we estimated the false-positive rates (FP) for each subtype group as

$$\mathrm{FP(SG)} = 1 - \frac{N^{T}_{\mathrm{WFC3}}(\mathrm{SG})}{N^{*}_{\mathrm{WFC3}}(\mathrm{SG})}, \tag{7}$$

where $N^{T}_{\mathrm{WFC3}}$ is the number of WFC3 spectra visually confirmed as UCDs. With the exception of the late-T and Y

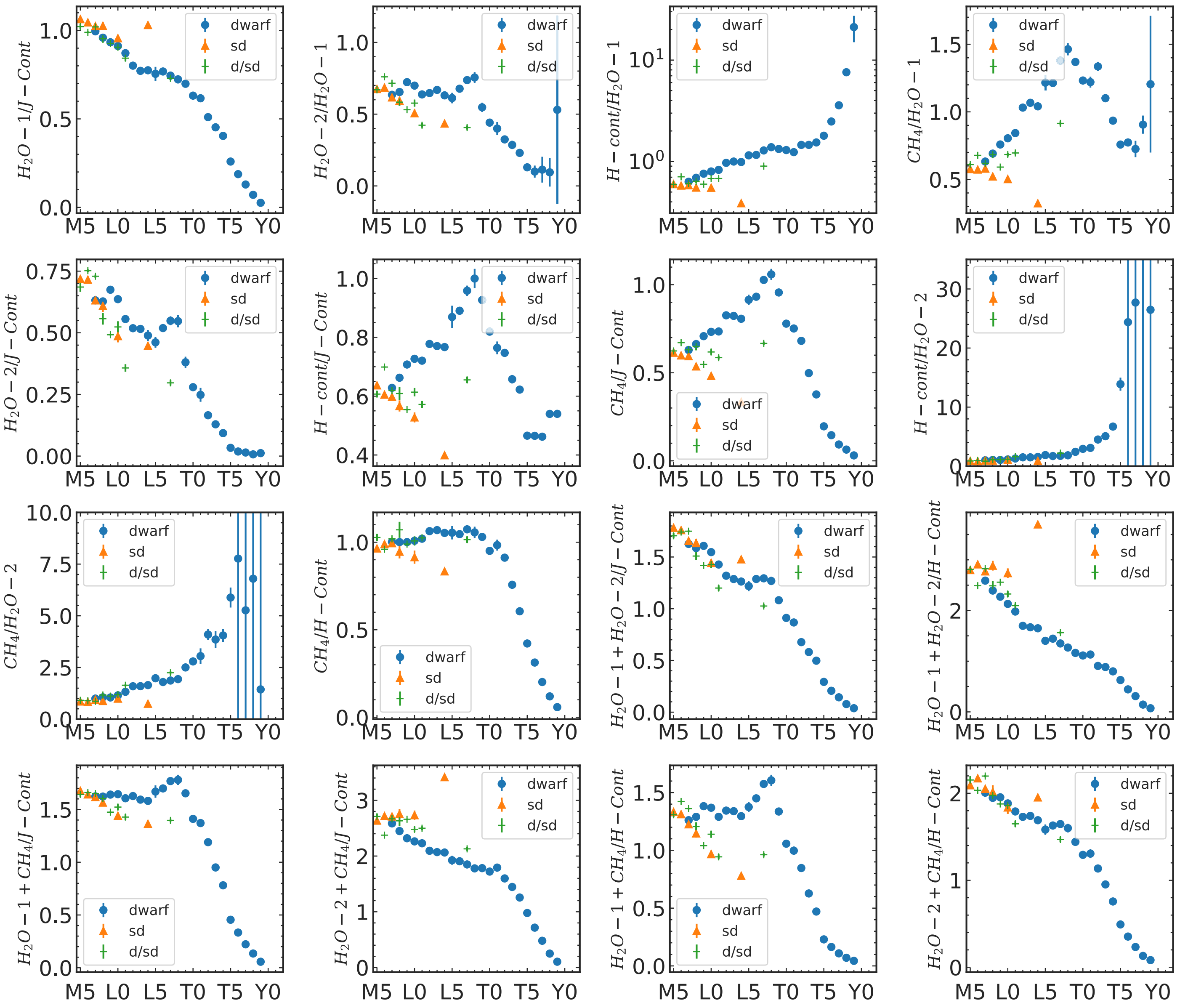

**Figure 4.** Spectral index values as a function of spectral type. Dwarf standards are indicated by blue dots, mild subdwarf (d/sd) standards by green crosses, and subdwarfs (sd) standards by orange triangles.

**Table 2**
Selection Criteria

| SpT Range | *X*-axis | *Y*-axis | *v*1 | *v*2 | *v*3 | *v*4 | CP | CT | FP |
|---|---|---|---|---|---|---|---|---|---|
| M7–M9 | $H_2O$-1/J-cont | $H_2O$-2+$CH_4$/H-cont | (0.83, 2.27) | (1.07, 2.49) | (1.07, 1.65) | (0.83, 1.42) | 0.99 | 0.164 | 0.890 |
| L0–L4 | $H_2O$-1/J-cont | $H_2O$-2+$CH_4$/H-cont | (0.68, 2.19) | (0.96, 2.61) | (0.96, 1.32) | (0.68, 0.9) | 1.0 | 0.070 | 0.903 |
| L5–L9 | $H_2O$-1+$H_2O$-2/H-cont | $H_2O$-1+$CH_4$/J-cont | (0.79, 2.84) | (1.81, 2.86) | (1.81, 0.6) | (0.79, 0.58) | 1.0 | 0.037 | 0.928 |
| T0–T4 | $H_2O$-2/J-cont | $H_2O$-1+$H_2O$-2/H-cont | (−0.1, 1.55) | (0.34, 2.5) | (0.34, 0.18) | (−0.1, -0.77) | 1.0 | 0.013 | 0.831 |
| T5–T9 | $H_2O$-1+$H_2O$-2/J-cont | $H_2O$-1+$CH_4$/H-cont | (−0.04, 0.31) | (0.36, 0.6) | (0.36, -0.03) | (−0.04, -0.32) | 1.0 | 0.0007 | 0.25 |
| Y dwarfs | $CH_4$/H-cont | $H_2O$-1+$H_2O$-2/J-cont | (−0.19, 0.63) | (0.32, 0.8) | (0.32, -0.63) | (−0.19, -0.8) | 0.95 | 0.0007 | 0.25 |
| Subdwarfs | $H_2O$-1+$CH_4$/J-cont | $H_2O$-2+$CH_4$/J-cont | (1.35, 3.82) | (1.86, 3.71) | (1.86, 1.6) | (1.35, 1.71) | 1.0 | 0.503 | 1.0 |

**Note.** v1-v4 denote the ($x$, $y$) positions of the vertices of the selection box for the given index-index pair; CP, CT, and FP are the completeness, contamination, and false-positive rates defined in Equations (5)–(7).

dwarfs, the best index selection criteria have FP values of nearly 100%, indicating significant contamination remains after these selection steps, although in terms of numbers these late-T and Y dwarfs select $< 15$ objects each. Nevertheless, the cumulative selection process reduces the number of spectra requiring visual inspection by over 99%, and spot checks of the rejected sample indicate that few if any UCDs were missed by this process.

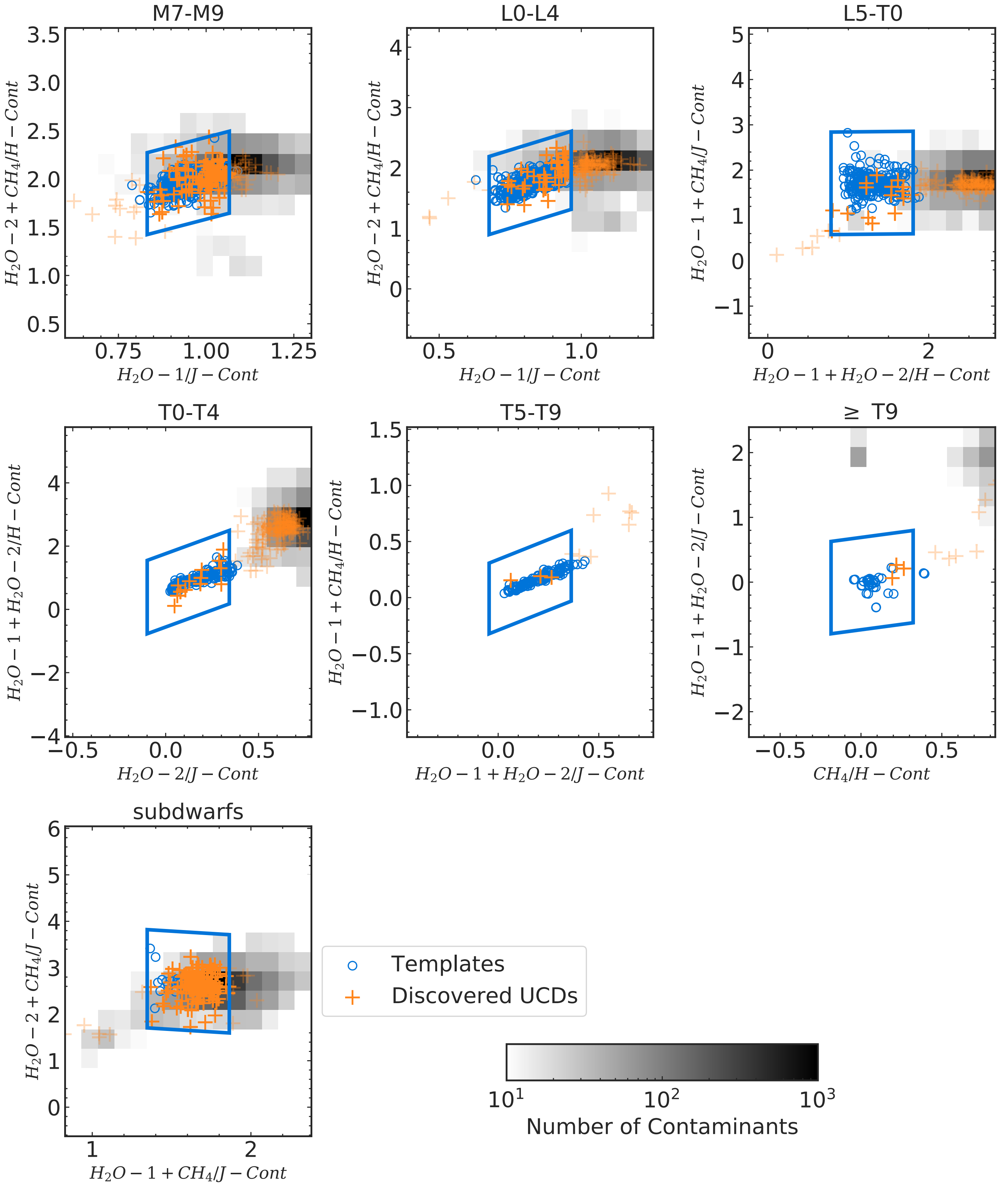

**Figure 5.** Optimal index-index selection regions for the subtype groups defined in this study. The shaded 2D histograms show the density of WISPS and 3D-HST point sources after applying both S/N and line-fit cuts. The blue circles are the appropriately classified UCD spectral templates used to define the selection regions. The orange pluses are all visually confirmed UCDs from the WISPS and 3D-HST samples.

**Table 3**
Late-M Dwarfs Identified in WISPS and 3D-HST Data

| Designation | Field ID | F110W | F140W | F160W | $J$ S/N | SpT | Distance(pc) | Selection[a] |
|---|---|---|---|---|---|---|---|---|
| J00114889-0653461 | PAR261-00027 | 21.135 ± 0.005 | ⋯ | 21.189 ± 0.011 | 25 | M7.0 ± 1.4 | $957^{+530}_{-254}$ | IND,DNN |
| J00483113-1247474 | PAR247-00043 | ⋯ | 22.167 ± 0.041 | ⋯ | 3 | M9.0 ± 4.2 | $840^{+2217}_{-416}$ | IND,RF |
| J01101162-0225003 | PAR83-00004 | 19.209 ± 0.001 | ⋯ | 18.624 ± 0.002 | 105 | M7.0 ± 9.3 | $361^{+5E7}_{-268}$ | IND |
| J01224506-2838463 | PAR128-00052 | ⋯ | ⋯ | 22.299 ± 0.021 | 14 | M8.0 ± 1.2 | $1365^{+429}_{-297}$ | IND,DNN |
| J01224822-2837534 | PAR128-00034 | ⋯ | ⋯ | 21.931 ± 0.013 | 16 | M8.0 ± 1.0 | $1174^{+339}_{-266}$ | IND,DNN,RF |
| J01371896-0908545 | PAR317-00032 | 21.678 ± 0.006 | ⋯ | 21.14 ± 0.009 | 17 | M8.0 ± 0.7 | $898^{+248}_{-199}$ | IND,DNN,RF |
| J02165968-0513381 | UDS-21-G141-14877 | ⋯ | 20.886 ± 0.004 | 20.832 ± 0.002 | 41 | M7.0 ± 0.5 | $719^{+221}_{-162}$ | IND,DNN |
| J02170221-0510136 | UDS-23-G141-31620 | ⋯ | 22.667 ± 0.016 | 22.727 ± 0.011 | 9 | M7.0 ± 2.0 | $1733^{+2185}_{-654}$ | IND,DNN |
| J02170328-0513353 | UDS-15-G141-15337 | ⋯ | 18.985 ± 0.001 | 18.978 ± 0.001 | 160 | M7.0 ± 0.5 | $302^{+98}_{-68}$ | IND,RF |
| J02171624-0513031 | UDS-09-G141-17647 | ⋯ | 21.617 ± 0.007 | 21.649 ± 0.004 | 21 | M7.0 ± 0.9 | $1041^{+417}_{-285}$ | IND,DNN |
| J02172178-0509209 | UDS-25-G141-36035 | ⋯ | 21.038 ± 0.004 | 21.029 ± 0.003 | 42 | M7.0 ± 0.5 | $782^{+241}_{-177}$ | IND,RF |
| J02172843-0514343 | UDS-10-G141-10211 | ⋯ | 20.75 ± 0.003 | 20.653 ± 0.002 | 46 | M8.0 ± 0.7 | $563^{+177}_{-130}$ | IND,DNN,RF |
| J02173221-0508120 | UDS-05-G141-41125 | ⋯ | 20.915 ± 0.003 | 20.933 ± 0.003 | 36 | M7.0 ± 0.6 | $744^{+252}_{-175}$ | IND,DNN |
| J02173525-0515329 | UDS-14-G141-05410 | ⋯ | 21.259 ± 0.005 | 21.226 ± 0.003 | 31 | M7.0 ± 0.5 | $860^{+235}_{-193}$ | IND,DNN,RF |
| J02174068-0514209 | UDS-14-G141-11264 | ⋯ | 21.683 ± 0.007 | 21.655 ± 0.004 | 21 | M7.0 ± 0.8 | $1076^{+368}_{-278}$ | IND,DNN,RF |
| J02182210-0509095 | UDS-18-G141-36798 | ⋯ | 21.867 ± 0.009 | 21.781 ± 0.008 | 16 | M7.0 ± 1.2 | $1152^{+587}_{-324}$ | IND,DNN |
| J03321669-2750086 | GOODSS-19-G141-16588 | ⋯ | 21.181 ± 0.004 | 21.131 ± 0.002 | 35 | M7.0 ± 0.5 | $828^{+254}_{-184}$ | IND,DNN,RF |
| J03321819-2747466 | GOODSS-02-G141-24465 | ⋯ | 22.282 ± 0.009 | 22.219 ± 0.004 | 13 | M8.0 ± 1.8 | $1190^{+710}_{-390}$ | IND,DNN,RF |
| J03321923-2745455 | GOODSS-09-G141-32414 | ⋯ | 22.274 ± 0.009 | 22.144 ± 0.005 | 17 | M8.0 ± 0.7 | $1139^{+320}_{-258}$ | IND,DNN |
| J03322070-2755018 | GOODSS-05-G141-01783 | ⋯ | 22.1 ± 0.008 | 22.109 ± 0.009 | 13 | M7.0 ± 1.2 | $1343^{+675}_{-407}$ | IND,DNN |
| J03322417-2742110 | GOODSS-30-G141-44380 | ⋯ | 20.124 ± 0.001 | 20.051 ± 0.001 | 74 | M7.0 ± 0.7 | $509^{+147}_{-119}$ | IND,DNN,RF |
| J03322960- 2752287 | GOODSS-13-G141-07509 | ⋯ | 22.045 ± 0.008 | 22.001 ± 0.007 | 16 | M8.0 ± 0.7 | $1048^{+333}_{-244}$ | IND,DNN |
| J03324208-2749116 | GOODSS-04-G141-19624 | ⋯ | 22.46 ± 0.012 | 22.392 ± 0.005 | 11 | M8.0 ± 1.7 | $1274^{+670}_{-396}$ | IND,DNN |
| J03324721-2744089 | GOODSS-29-G141-37362 | ⋯ | 20.03 ± 0.001 | 20.031 ± 0.001 | 86 | M7.0 ± 0.6 | $494^{+158}_{-117}$ | IND,RF |
| J03325279-2751257 | GOODSS-06-G141-11322 | ⋯ | 23.153 ± 0.021 | 23.031 ± 0.015 | 7 | M7.0 ± 1.7 | $2029^{+1891}_{-614}$ | IND,DNN |
| J03325579-2751454 | GOODSS-06-G141-10354 | ⋯ | 20.459 ± 0.002 | 20.423 ± 0.003 | 57 | M7.0 ± 0.6 | $598^{+181}_{-136}$ | IND,DNN,RF |
| J03330297-2751143 | GOODSS-28-G141-12490 | ⋯ | 21.992 ± 0.007 | 22.155 ± 0.005 | 18 | M7.0 ± 0.6 | $1252^{+438}_{-319}$ | IND,DNN |
| J03330928-2751045 | GOODSS-28-G141-12948 | ⋯ | 21.695 ± 0.006 | 21.857 ± 0.004 | 22 | M7.0 ± 0.6 | $1100^{+443}_{-289}$ | IND,DNN,RF |
| J05021436+0732089 | PAR189-00077 | ⋯ | 22.247 ± 0.016 | ⋯ | 7 | M9.0 ± 1.5 | $850^{+346}_{-224}$ | IND,DNN |
| J08391530+6456568 | PAR250-00051 | ⋯ | 22.327 ± 0.023 | ⋯ | 7 | M7.0 ± 1.5 | $1269^{+911}_{-386}$ | IND,DNN |
| J08540020+4351119 | PAR319-00085 | 23.212 ± 0.015 | ⋯ | 22.876 ± 0.025 | 7 | M8.0 ± 1.2 | $1937^{+663}_{-458}$ | IND |
| J09081156+3246358 | PAR417-00014 | 20.759 ± 0.004 | ⋯ | 20.277 ± 0.006 | 30 | M7.0 ± 0.9 | $714^{+255}_{-183}$ | IND,DNN,RF |
| J09102912+3328008 | PAR431- 00028 | 21.492 ± 0.006 | ⋯ | 21.323 ± 0.011 | 11 | M8.0 ± 0.9 | $905^{+225}_{-181}$ | DNN |
| J09110260+3331437 | PAR472-00057 | ⋯ | ⋯ | 22.107 ± 0.016 | 13 | M7.0 ± 4.4 | $1486^{+1.2E5}_{-692}$ | IND,DNN |
| J09113305+1832308 | PAR271-00055 | 22.242 ± 0.008 | ⋯ | 21.825 ± 0.013 | 10 | M7.0 ± 1.5 | $1432^{+893}_{-423}$ | IND,DNN,RF |
| J09260832+1239515 | PAR92-00011 | ⋯ | ⋯ | 20.725 ± 0.008 | 27 | M7.0 ± 0.8 | $773^{+245}_{-175}$ | IND,RF |
| J09443679-1941456 | PAR293-00059 | ⋯ | 21.95 ± 0.016 | ⋯ | 8 | M7.0 ± 1.6 | $1042^{+781}_{-326}$ | DNN |
| J09443831-1940444 | PAR293-00045 | ⋯ | 21.51 ± 0.012 | ⋯ | 13 | M7.0 ± 1.5 | $867^{+646}_{-263}$ | IND,DNN |
| J09470000+5126334 | PAR478-00038 | 22.253 ± 0.009 | ⋯ | 21.633 ± 0.012 | 17 | M7.0 ± 0.9 | $1386^{+535}_{-383}$ | IND,DNN,RF |
| J09505808+3544046 | PAR192-00026 | ⋯ | 21.467 ± 0.01 | ⋯ | 18 | M7.0 ± 0.6 | $842^{+243}_{-166}$ | IND,DNN,RF |
| J09590957+5549134 | PAR246-00017 | ⋯ | 20.91 ± 0.007 | ⋯ | 21 | M7.0 ± 1.6 | $649^{+463}_{-182}$ | IND,DNN,RF |
| J09591621+5549034 | PAR246-00056 | ⋯ | 22.614 ± 0.021 | ⋯ | 5 | M7.0 ± 3.3 | $1424^{+6704}_{-691}$ | IND, |

**Table 3**
(Continued)

| Designation | Field ID | F110W | F140W | F160W | $J$ S/N | SpT | Distance(pc) | Selection[a] |
|---|---|---|---|---|---|---|---|---|
| J10001957+0218224 | COSMOS-26-G141-12464 | ⋯ | 21.682 ± 0.005 | 21.659 ± 0.004 | 16 | M7.0 ± 1.2 | $1086^{+525}_{-319}$ | IND,DNN,RF |
| J10002575+0222265 | COSMOS-25-G141-19163 | ⋯ | 22.516 ± 0.01 | 22.343 ± 0.007 | 9 | M8.0 ± 2.4 | $1285^{+1160}_{-438}$ | IND,DNN,RF |
| J10002739+0212135 | COSMOS-14-G141-02407 | ⋯ | 21.408 ± 0.004 | 21.426 ± 0.005 | 23 | M7.0 ± 0.7 | $943^{+337}_{-232}$ | IND,DNN,RF |
| J10003031+0227345 | COSMOS-08-G141-26927 | ⋯ | 21.988 ± 0.007 | 21.874 ± 0.007 | 13 | M7.0 ± 1.6 | $1209^{+952}_{-378}$ | IND,DNN |
| J10052069-2421432 | PAR336-00047 | 22.108 ± 0.006 | ⋯ | 22.041 ± 0.012 | 10 | M8.0 ± 1.1 | $1224^{+386}_{-257}$ | IND,DNN,RF |
| J10052583+0130297 | PAR104-00019 | 21.552 ± 0.003 | ⋯ | 21.896 ± 0.01 | 27 | M7.0 ± 1.2 | $1247^{+547}_{-320}$ | IND,DNN |
| J10053689-2304480 | PAR349-00079 | 22.665 ± 0.123 | ⋯ | 22.472 ± 0.229 | 8 | M7.0 ± 0.9 | $1822^{+636}_{-418}$ | IND,DNN,RF |
| J10065538-2953384 | PAR170-00081 | ⋯ | ⋯ | 21.903 ± 0.017 | 13 | M8.0 ± 1.0 | $1137^{+379}_{-244}$ | IND |
| J10074034+1004444 | PAR343-00036 | 21.476 ± 0.006 | ⋯ | 21.234 ± 0.01 | 15 | M8.0 ± 0.8 | $889^{+230}_{-189}$ | IND,DNN,RF |
| J10074043+5012192 | PAR98-00082 | ⋯ | ⋯ | 22.698 ± 0.033 | 7 | M7.0 ± 2.1 | $1935^{+1947}_{-615}$ | IND,DNN,RF |
| J10074625+5013373 | PAR98-00038 | ⋯ | ⋯ | 21.474 ± 0.012 | 21 | M9.0 ± 1.7 | $843^{+293}_{-205}$ | IND,DNN,RF |
| J10093821+3000442 | PAR39-00033 | ⋯ | 22.062 ± 0.016 | ⋯ | 10 | M7.0 ± 1.3 | $1139^{+553}_{-296}$ | IND,DNN |
| J10232252+0409233 | PAR347-00017 | 20.691 ± 0.002 | ⋯ | 20.47 ± 0.004 | 37 | M7.0 ± 1.0 | $736^{+271}_{-172}$ | IND,DNN,RF |
| J10244408-1843228 | PAR179-00069 | ⋯ | 21.992 ± 0.014 | ⋯ | 13 | M7.0 ± 1.4 | $1075^{+643}_{-302}$ | IND,DNN |
| J11021593+1053514 | PAR11-00046 | 22.302 ± 0.01 | 22.221 ± 0.014 | ⋯ | 11 | M7.0 ± 0.6 | $1410^{+418}_{-365}$ | IND,DNN,RF |
| J11121413+3536286 | PAR44-00044 | ⋯ | 21.573 ± 0.009 | ⋯ | 19 | M8.0 ± 1.0 | $742^{+237}_{-172}$ | IND,DNN,RF |
| J11252211+5319527 | PAR477-00009 | 20.799 ± 0.005 | ⋯ | 20.435 ± 0.008 | 32 | M8.0 ± 0.7 | $627^{+154}_{-129}$ | IND,DNN,RF |
| J12242636+6110570 | PAR422-00021 | ⋯ | 21.353 ± 0.005 | ⋯ | 30 | M8.0 ± 1.4 | $688^{+322}_{-179}$ | IND,DNN,RF |
| J12361834+6214264 | GOODSN-13-G141-20147 | ⋯ | 21.771 ± 0.003 | 21.727 ± 0.004 | 29 | M8.0 ± 0.7 | $925^{+280}_{-223}$ | IND,DNN |
| J12365687+6210019 | GOODSN-43-G141-05338 | ⋯ | 19.962 ± 0.001 | 19.929 ± 0.001 | 88 | M7.0 ± 0.5 | $475^{+138}_{-108}$ | IND,DNN |
| J12365813+6218512 | GOODSN-16-G141-33587 | ⋯ | 21.85 ± 0.005 | 21.809 ± 0.005 | 22 | M8.0 ± 0.7 | $955^{+306}_{-230}$ | IND,DNN |
| J12370991+6210081 | GOODSN-43-G141-05553 | ⋯ | 20.273 ± 0.001 | 20.302 ± 0.001 | 78 | M7.0 ± 0.5 | $557^{+177}_{-130}$ | IND,DNN,RF |
| J12371407+6219051 | GOODSN-27-G141-34168 | ⋯ | 21.789 ± 0.005 | 21.785 ± 0.004 | 20 | M7.0 ± 1.0 | $1116^{+487}_{-297}$ | IND,DNN |
| J12423940+3538175 | PAR439-00106 | 24.076 ± 0.029 | ⋯ | 23.353 ± 0.034 | 4 | M7.0 ± 2.0 | $3145^{+3440}_{-1089}$ | IND,DNN |
| J12565999+5430547 | PAR110-00085 | ⋯ | ⋯ | 22.36 ± 0.025 | 15 | M8.0 ± 1.3 | $1395^{+536}_{-319}$ | IND,DNN |
| J13034862+2952039 | PAR35- 00023 | ⋯ | 20.782 ± 0.005 | ⋯ | 43 | M8.0 ± 0.7 | $507^{+125}_{-92}$ | IND,DNN |
| J13051928-2538175 | PAR32-00044 | ⋯ | 21.986 ± 0.014 | ⋯ | 17 | M8.0 ± 0.8 | $879^{+254}_{-178}$ | IND |
| J13193620+2727034 | PAR47-00010 | ⋯ | 20.545 ± 0.004 | ⋯ | 39 | M7.0 ± 1.9 | $564^{+675}_{-198}$ | IND,DNN,RF |
| J13234291+3434440 | PAR186-00091 | ⋯ | 22.233 ± 0.029 | ⋯ | 6 | M8.0 ± 2.5 | $990^{+1225}_{-371}$ | IND, |
| J13253029+2233198 | PAR436-00037 | 21.778 ± 0.019 | ⋯ | 21.23 ± 0.026 | 20 | M7.0 ± 0.5 | $1119^{+298}_{-261}$ | IND,DNN,RF |
| J13302849+2810315 | PAR52-00082 | ⋯ | 23.2 ± 0.028 | ⋯ | 7 | M8.0 ± 2.1 | $1562^{+1178}_{-516}$ | IND |
| J13405721+2825163 | PAR433-00047 | 21.903 ± 0.008 | ⋯ | 22.211 ± 0.023 | 14 | M8.0 ± 0.8 | $1204^{+331}_{-246}$ | IND,DNN |
| J13422570+1841485 | PAR139-00010 | ⋯ | ⋯ | 20.226 ± 0.005 | 70 | M7.0 ± 0.6 | $607^{+163}_{-124}$ | IND,DNN,RF |
| J13481140+2451529 | PAR243-00025 | ⋯ | 21.706 ± 0.013 | ⋯ | 10 | M8.0 ± 1.7 | $780^{+451}_{-232}$ | IND,DNN |
| J13510084+2751086 | PAR444-00034 | 21.762 ± 0.006 | ⋯ | 21.472 ± 0.012 | 16 | M9.0 ± 1.1 | $881^{+241}_{-187}$ | IND,DNN,RF |
| J13531331+3619253 | PAR350-00043 | 22.324 ± 0.007 | ⋯ | 22.1 ± 0.014 | 13 | M7.0 ± 1.1 | $1544^{+621}_{-375}$ | IND,DNN |
| J14022476+0946090 | PAR143-00045 | 22.097 ± 0.007 | ⋯ | 21.862 ± 0.012 | 11 | M8.0 ± 0.8 | $1168^{+292}_{-219}$ | IND |
| J14024558+5410246 | PAR458- 00004 | 19.061 ± 0.001 | ⋯ | 18.889 ± 0.003 | 107 | M8.0 ± 0.7 | $295^{+71}_{-55}$ | IND,DNN |
| J14081989+5657245 | PAR353-00055 | 22.498 ± 0.013 | ⋯ | 21.994 ± 0.018 | 13 | M9.0 ± 1.6 | $1160^{+443}_{-280}$ | IND,DNN,RF |
| J14101216+2954504 | PAR222-00091 | ⋯ | 23.149 ± 0.03 | ⋯ | 4 | M9.0 ± 3.2 | $1263^{+1632}_{-538}$ | IND,RF |
| J14101279+2955426 | PAR222-00047 | ⋯ | 22.12 ± 0.016 | ⋯ | 8 | M7.0 ± 1.4 | $1159^{+833}_{-336}$ | IND,DNN |
| J14185811+5244430 | AEGIS-25-G141-18460 | ⋯ | 22.792 ± 0.012 | 22.822 ± 0.012 | 10 | M7.0 ± 0.9 | $1808^{+789}_{-484}$ | IND,DNN,RF |

**Table 3**
(Continued)

| Designation | Field ID | F110W | F140W | F160W | $J$ S/N | SpT | Distance(pc) | Selection[a] |
|---|---|---|---|---|---|---|---|---|
| J14191922+5253001 | AEGIS-11-G141-37605 | ⋯ | 21.75 ± 0.005 | 21.663 ± 0.004 | 23 | M7.0 ± 0.9 | $1079^{+411}_{-270}$ | IND,DNN |
| J14195553+5253327 | AEGIS-06-G141-12749 | ⋯ | 22.589 ± 0.007 | 22.488 ± 0.007 | 10 | M7.0 ± 1.2 | $1598^{+798}_{-456}$ | IND,DNN,RF |
| J14201199+5254145 | AEGIS-07-G141-04371 | ⋯ | 21.377 ± 0.003 | 21.343 ± 0.003 | 31 | M7.0 ± 0.5 | $908^{+282}_{-201}$ | IND,DNN,RF |
| J14271276+2631084 | PAR218-00032 | ⋯ | 21.413 ± 0.015 | ⋯ | 13 | M7.0 ± 1.0 | $835^{+337}_{-202}$ | IND,DNN |
| J14291997+3224590 | PAR378-00052 | 22.432 ± 0.011 | ⋯ | 21.922 ± 0.016 | 8 | M9.0 ± 1.8 | $1152^{+446}_{-319}$ | IND,DNN,RF |
| J14312496+2447492 | PAR385-00049 | 22.841 ± 0.015 | ⋯ | 22.191 ± 0.019 | 11 | M7.0 ± 1.1 | $1795^{+752}_{-538}$ | IND |
| J14320077+0958067 | PAR428-00062 | 22.212 ± 0.012 | ⋯ | 21.976 ± 0.022 | 9 | M8.0 ± 1.2 | $1240^{+445}_{-296}$ | IND,DNN,RF |
| J14372742-0149429 | PAR66-00029 | 21.372 ± 0.004 | ⋯ | 21.584 ± 0.01 | 20 | M7.0 ± 0.7 | $1096^{+301}_{-227}$ | IND,RF |
| J15001904+4127259 | PAR391-00011 | 20.929 ± 0.002 | ⋯ | 20.719 ± 0.005 | 33 | M8.0 ± 0.7 | $692^{+172}_{-137}$ | IND,DNN,RF |
| J15140761+3617309 | PAR71-00034 | 22.091 ± 0.006 | ⋯ | 21.894 ± 0.012 | 10 | M8.0 ± 0.7 | $1187^{+275}_{-231}$ | IND |
| J15141175+3616405 | PAR71-00038 | 22.154 ± 0.008 | ⋯ | 21.595 ± 0.011 | 13 | M7.0 ± 1.5 | $1368^{+835}_{-432}$ | IND,DNN,RF |
| J15345595+1252516 | PAR457-00025 | 21.507 ± 0.006 | ⋯ | 21.563 ± 0.011 | 16 | M7.0 ± 0.7 | $1109^{+318}_{-221}$ | IND,DNN,RF |
| J15400156-0204397 | PAR446-00065 | 22.198 ± 0.006 | ⋯ | 21.696 ± 0.009 | 16 | M7.0 ± 1.1 | $1372^{+558}_{-368}$ | IND,DNN |
| J15425688-1051180 | PAR449-00079 | 22.493 ± 0.008 | ⋯ | 22.324 ± 0.015 | 8 | M7.0 ± 1.2 | $1707^{+862}_{-416}$ | IND,DNN |
| J15451481+1155008 | PAR290-00009 | ⋯ | 19.164 ± 0.003 | ⋯ | 76 | M7.0 ± 0.7 | $295^{+83}_{-63}$ | IND |
| J15502299+3959298 | PAR59-00072 | ⋯ | 22.752 ± 0.019 | ⋯ | 13 | M7.0 ± 1.1 | $1506^{+654}_{-366}$ | IND,DNN |
| J15563316+2107548 | PAR308-00020 | 21.283 ± 0.004 | ⋯ | 21.132 ± 0.008 | 12 | M7.0 ± 1.1 | $965^{+424}_{-219}$ | IND,DNN,RF |
| J16045847+1445598 | PAR240-00058 | ⋯ | 22.411 ± 0.017 | ⋯ | 8 | M8.0 ± 1.2 | $1082^{+397}_{-256}$ | IND,DNN |
| J16052496+2547388 | PAR148- 00044 | ⋯ | ⋯ | 21.913 ± 0.016 | 21 | M7.0 ± 1.1 | $1347^{+585}_{-321}$ | IND,DNN |
| J17031213+6136515 | PAR155-00040 | ⋯ | ⋯ | 21.64 ± 0.013 | 22 | M7.0 ± 0.8 | $1189^{+404}_{-281}$ | IND,RF |
| J17150193+0455305 | PAR239-00118 | ⋯ | 22.097 ± 0.018 | ⋯ | 6 | M9.0 ± 2.2 | $791^{+442}_{-263}$ | IND,DNN,RF |
| J17200596+4805092 | PAR398-00028 | 21.265 ± 0.003 | ⋯ | 20.801 ± 0.005 | 32 | M7.0 ± 0.5 | $893^{+249}_{-204}$ | IND |
| J17391964+4554544 | PAR382-00026 | 20.854 ± 0.003 | ⋯ | 20.274 ± 0.004 | 45 | M7.0 ± 1.6 | $731^{+531}_{-220}$ | DNN,RF |
| J18323319+5345108 | PAR124-00065 | 22.139 ± 0.01 | ⋯ | 22.097 ± 0.02 | 13 | M7.0 ± 1.1 | $1478^{+631}_{-360}$ | IND,RF |
| J20054131-4139000 | PAR371-00080 | 22.214 ± 0.007 | ⋯ | 21.839 ± 0.012 | 10 | M7.0 ± 1.2 | $1428^{+694}_{-366}$ | IND,DNN |
| J20054469-4139216 | PAR371-00045 | 21.308 ± 0.004 | ⋯ | 21.189 ± 0.009 | 19 | M7.0 ± 0.6 | $986^{+243}_{-184}$ | IND,DNN,RF |
| J20220176-3113120 | PAR332-00100 | 22.119 ± 0.009 | ⋯ | 21.975 ± 0.017 | 13 | M8.0 ± 0.7 | $1206^{+299}_{-229}$ | IND,DNN |
| J20382223-2021468 | PAR197-00054 | ⋯ | 21.101 ± 0.012 | ⋯ | 12 | M7.0 ± 1.3 | $732^{+362}_{-222}$ | IND,DNN,RF |
| J20402638-0644161 | PAR248-00079 | ⋯ | 21.627 ± 0.011 | ⋯ | 17 | M8.0 ± 0.7 | $750^{+190}_{-150}$ | IND,DNN |
| J21040517-0723376 | PAR43-00060 | 21.949 ± 0.006 | 21.833 ± 0.009 | ⋯ | 13 | M7.0 ± 0.7 | $1173^{+379}_{-309}$ | IND |
| J21391100- 3824072 | PAR309-00046 | 22.323 ± 0.008 | ⋯ | 22.355 ± 0.017 | 9 | M8.0 ± 1.2 | $1397^{+468}_{-320}$ | IND |
| J21391186-3824108 | PAR309-00023 | 21.186 ± 0.004 | ⋯ | 20.879 ± 0.008 | 29 | M8.0 ± 0.7 | $763^{+178}_{-156}$ | IND,DNN,RF |
| J22222288+0937084 | PAR50-00007 | ⋯ | 19.715 ± 0.003 | ⋯ | 64 | M7.0 ± 0.7 | $382^{+104}_{-79}$ | IND |
| J22253721-7212288 | PAR404-00044 | 21.885 ± 0.005 | ⋯ | 21.715 ± 0.01 | 15 | M8.0 ± 1.0 | $1080^{+297}_{-218}$ | IND |
| J22484228-8011132 | PAR372-00114 | 23.001 ± 0.01 | ⋯ | 22.479 ± 0.015 | 9 | M7.0 ± 1.0 | $1998^{+709}_{-500}$ | IND,DNN |
| J23071667+2112090 | PAR166-00044 | ⋯ | ⋯ | 21.599 ± 0.013 | 15 | M8.0 ± 1.2 | $999^{+343}_{-240}$ | IND |
| J23071720+2112302 | PAR166-00041 | ⋯ | ⋯ | 21.529 ± 0.018 | 17 | M7.0 ± 1.3 | $1128^{+488}_{-276}$ | IND |
| J23333572+3922141 | PAR68-00017 | 18.685 ± 0.001 | ⋯ | 18.697 ± 0.002 | 146 | M8.0 ± 0.6 | $258^{+61}_{-45}$ | IND,DNN,RF |
| J23333806+3921333 | PAR68-00027 | 20.521 ± 0.002 | ⋯ | 20.307 ± 0.005 | 38 | M8.0 ± 0.9 | $574^{+155}_{-116}$ | IND,DNN |
| J23333951+3925052 | PAR153-00002 | ⋯ | ⋯ | 15.74 ± 0.0 | 518 | M8.0 ± 1.1 | $68^{+21}_{-15}$ | IND,DNN,RF |
| J23333997+3924558 | PAR153-00048 | ⋯ | ⋯ | 21.537 ± 0.01 | 30 | M7.0 ± 1.7 | $1170^{+919}_{-350}$ | IND,DNN,RF |
| J23351983-3536072 | PAR359-00007 | 19.871 ± 0.002 | ⋯ | 19.593 ± 0.003 | 58 | M7.0 ± 0.6 | $487^{+129}_{-103}$ | IND,DNN,RF |

**Table 3**
(Continued)

| Designation | Field ID | F110W | F140W | F160W | $J$ S/N | SpT | Distance(pc) | Selection[a] |
|---|---|---|---|---|---|---|---|---|
| J23450002+1510350 | PAR77- 00045 | ⋯ | ⋯ | $21.683 \pm 0.015$ | 12 | M8.0 $\pm$ 1.0 | $1020^{+307}_{-226}$ | IND,DNN |
| J23450092-4239288 | PAR356-00057 | $23.028 \pm 0.013$ | ⋯ | $22.657 \pm 0.022$ | 8 | M7.0 $\pm$ 1.3 | $2108^{+1148}_{-588}$ | IND,DNN,RF |

**Note.** Table 3 is available in machine-readable format.
[a] Selection methods are index-index selection (IND), random forest selection (RF), and deep neural network (DNN).

(This table is available in machine-readable form.)

**Table 4**
L and T Dwarfs Identified in WISPS and 3D-HST Data

| Designation | Field ID | F110W | F140W | F160W | *J* S/N | SpT | Distance (pc) | Selection[a] |
|---|---|---|---|---|---|---|---|---|
| J00150859-7955488 | PAR244-00072 | ⋯ | 22.218 ± 0.016 | ⋯ | 6 | L2.0 ± 2.0 | $512^{+240}_{-163}$ | IND |
| J02170016-0509564 | UDS-23-G141-32939 | ⋯ | 23.864 ± 0.05 | 23.706 ± 0.023 | 4 | L6.0 ± 2.7 | $802^{+478}_{-267}$ | IND,DNN,RF |
| J02171640-0509133 | UDS-25-G141-36758 | ⋯ | 21.318 ± 0.005 | 21.04 ± 0.005 | 31 | L1.0 ± 0.6 | $476^{+161}_{-115}$ | IND,DNN,RF |
| J02465310-0104453 | PAR483-00077 | 23.002 ± 0.012 | ⋯ | 22.089 ± 0.013 | 9 | L1.0 ± 1.1 | $1065^{+354}_{-294}$ | IND,DNN,RF |
| J03074119-7243574 | PAR130-00092 | ⋯ | ⋯ | 22.726 ± 0.036 | 12 | T4.0 ± 0.5 | $322^{+83}_{-68}$ | IND,DNN,RF |
| J03260271-1643210 | PAR467-00135 | 23.862 ± 0.043 | ⋯ | 23.925 ± 0.098 | 3 | T1.0 ± 1.5 | $739^{+180}_{-126}$ | IND,DNN,RF |
| J03322479-2749129 | GOODSS-20-G141-19648 | ⋯ | 23.244 ± 0.022 | 23.151 ± 0.008 | 6 | L1.0 ± 2.8 | $1236^{+722}_{-520}$ | IND,DNN |
| J03323881-2749536 | GOODSS-04-G141-17402 | ⋯ | 22.552 ± 0.019 | 22.942 ± 0.018 | 13 | T4.0 ± 0.9 | $296^{+121}_{-75}$ | IND,RF |
| J03325821-2741436 | GOODSS-01-G141-45889 | ⋯ | 22.149 ± 0.008 | 22.935 ± 0.01 | 31 | T6.0 ± 0.5 | $165^{+69}_{-48}$ | IND,DNN,RF |
| J03330420-2751369 | GOODSS-28-G141-10859 | ⋯ | 21.353 ± 0.004 | 21.325 ± 0.003 | 34 | L1.0 ± 0.7 | $512^{+198}_{-142}$ | IND,DNN,RF |
| J04375775- 1106158 | PAR463-00176 | 24.331 ± 0.054 | ⋯ | 24.285 ± 0.113 | 4 | T4.0 ± 0.9 | $723^{+180}_{-176}$ | IND,DNN,RF |
| J09275744+6027467 | PAR21-00005 | ⋯ | 16.697 ± 0.001 | ⋯ | 324 | L1.0 ± 0.7 | $49^{+13}_{-10}$ | IND,DNN,RF |
| J10001529+0221196 | COSMOS-27-G141-17304 | ⋯ | 24.401 ± 0.063 | 24.742 ± 0.051 | 5 | L4.0 ± 4.0 | $1523^{+1522}_{-745}$ | IND |
| J10002236+0219530 | COSMOS-03-G141-14879 | ⋯ | 23.233 ± 0.022 | 23.04 ± 0.017 | 6 | L4.0 ± 3.8 | $786^{+660}_{-331}$ | IND |
| J10003503+0217012 | COSMOS-23-G141-10232 | ⋯ | 23.844 ± 0.033 | 23.776 ± 0.022 | 5 | T0.0 ± 3.6 | $609^{+335}_{-183}$ | IND,RF |
| J10004273+0220589 | COSMOS-09-G141-16730 | ⋯ | 24.511 ± 0.069 | 24.956 ± 0.056 | 4 | L3.0 ± 6.3 | $1968^{+3595}_{-1168}$ | IND |
| J10034053+2854461 | PAR191-00077 | ⋯ | ⋯ | 23.067 ± 0.042 | 6 | L9.0 ± 3.7 | $611^{+299}_{-199}$ | IND |
| J10193326+2743134 | PAR201-00044 | ⋯ | 22.411 ± 0.029 | ⋯ | 4 | L6.0 ± 4.0 | $327^{+297}_{-84}$ | IND,DNN,RF |
| J10232247+0409534 | PAR347-00037 | 22.174 ± 0.007 | ⋯ | 21.953 ± 0.013 | 14 | L0.0 ± 1.3 | $958^{+277}_{-217}$ | IND,DNN,RF |
| J11151423+5257050 | PAR468-00163 | 24.256 ± 0.04 | ⋯ | 24.397 ± 0.101 | 5 | T2.0 ± 2.2 | $842^{+230}_{-216}$ | IND,DNN,RF |
| J11240834+4202344 | PAR106-00047 | ⋯ | ⋯ | 21.475 ± 0.013 | 11 | L7.0 ± 1.4 | $327^{+88}_{-69}$ | IND,DNN,RF |
| J11330584+0328395 | PAR27-00036 | 22.388 ± 0.011 | 22.182 ± 0.014 | ⋯ | 10 | L2.0 ± 1.2 | $656^{+270}_{-216}$ | IND,DNN,RF |
| J11415016+2640361 | PAR219-00083 | ⋯ | 23.261 ± 0.041 | ⋯ | 3 | L3.0 ± 3.7 | $714^{+692}_{-280}$ | IND |
| J11504964-2033396 | PAR199-00009 | ⋯ | 19.195 ± 0.003 | ⋯ | 57 | L1.0 ± 0.9 | $153^{+41}_{-31}$ | IND,DNN,RF |
| J11545284+1939360 | PAR338-00136 | 24.067 ± 0.046 | ⋯ | 23.06 ± 0.042 | 4 | L4.0 ± 2.9 | $1107^{+679}_{-375}$ | IND,RF |
| J12324241-0033067 | PAR58-00112 | ⋯ | 23.142 ± 0.045 | ⋯ | 11 | T7.0 ± 0.9 | $148^{+86}_{-65}$ | IND,DNN,RF |
| J12355232+6211418 | GOODSN-11-G141-10603 | ⋯ | 23.32 ± 0.016 | 23.043 ± 0.014 | 6 | L6.0 ± 2.3 | $602^{+310}_{-174}$ | IND,DNN,RF |
| J12361989+6209339 | GOODSN-31-G141-04491 | ⋯ | 24.253 ± 0.035 | 24.107 ± 0.036 | 5 | L3.0 ± 3.7 | $1442^{+1208}_{-649}$ | IND,DNN |
| J12363821+6209511 | GOODSN-32-G141-05180 | ⋯ | 24.221 ± 0.048 | 24.113 ± 0.04 | 4 | L3.0 ± 3.8 | $1435^{+1266}_{-671}$ | IND,DNN |
| J12363885+6214516 | GOODSN-24-G141-21552 | ⋯ | 22.035 ± 0.005 | 21.835 ± 0.004 | 19 | L4.0 ± 0.8 | $430^{+185}_{-124}$ | IND,DNN,RF |
| J12365374+6211177 | GOODSN-33-G141-09283 | ⋯ | 22.218 ± 0.007 | 21.969 ± 0.004 | 12 | L4.0 ± 1.8 | $477^{+254}_{-160}$ | IND,DNN,RF |
| J13052550-2538287 | PAR32-00075 | ⋯ | 23.045 ± 0.028 | ⋯ | 11 | T8.0 ± 0.7 | $87^{+53}_{-37}$ | IND,DNN,RF |
| J14185040+5242593 | AEGIS-03-G141-17053 | ⋯ | 22.706 ± 0.011 | 23.116 ± 0.014 | 21 | T4.0 ± 0.7 | $319^{+127}_{-75}$ | IND,DNN,RF |
| J16050161+1447002 | PAR240- 00040 | ⋯ | 21.957 ± 0.013 | ⋯ | 17 | L0.0 ± 1.6 | $654^{+250}_{-188}$ | IND,DNN,RF |
| J16184979+3340175 | PAR65-00035 | 21.659 ± 0.004 | ⋯ | 21.313 ± 0.009 | 19 | L0.0 ± 0.7 | $726^{+171}_{-135}$ | IND,DNN,RF |
| J16252493+5721274 | PAR156-00041 | ⋯ | ⋯ | 21.381 ± 0.011 | 19 | L4.0 ± 0.8 | $438^{+115}_{-91}$ | IND,DNN,RF |

**Note.** Table 4 is available in machine-readable format.
[a] Selection methods are index-index selection (IND), random forest selection (RF), and deep neural network (DNN).

(This table is available in machine-readable form.)

**Table 5**
Random Forest Model Parameters

| Parameter[a] | Description | Best Value |
|---|---|---|
| `class_weight` | Weight of each class | Balanced |
| `bootstrap` | Using bootstrap samples[b] | True |
| `oob_score` | Using out-of-the-bag samples to compute accuracy[b] | False |
| `n_estimators` | Number of trees | 947 |
| `min_impurity_split` | Minimum value of classification error to split the tree node on | 1.2E-8 |
| `max_leaf_nodes` | The maximum number of nodes for each leaf | 526 |
| `max_depth` | Maximum depth of a tree | 889 |
| `min_samples_split` | Minimum number of objects needed to split on an internal node | 53 |
| `criterion` | Measure of the quality of the split (node purity) on each tree | Entropy[c] |

**Notes.**
[a] Parameter name in the scikitlearn `RandomForestClassifier` object.
[b] Bootstrapping was done by splitting the training set into two randomly-assigned 50% partitions, training the random forest on one partition, and using this to classify the second partition (see, for example, Miller et al. 2017). Out-of-the-bag (OOB) samples are the data that were not used when fitting a tree to the bootstrapped samples
[c] The entropy is defined as $-\sum_{k=1}^{K} p_{\rm mk} \log p_{\rm mk}$, where $p_{mk}$ is the proportion of $k$-class objects classified $m$ class, where $K = 5$ is the total number of classes.

### *3.4. Selection by Random Forest Classifier*

Despite a high completeness and a significant reduction in the contaminants with the spectral index selection, this method selects more than 3000 candidates for visual inspection. To further reduce this number, we explored a different selection: a random forest classifier. A random forest (Breiman 2001) is a set of decision trees constructed on randomized samples of a data set. A decision tree is a model that recursively splits the sample into two parts until a predefined stopping criterion is reached. For example, if values of a feature $j$ are $\{X_j\}$, the decision tree could split this data set into two parts, $\{X|X_j < s\}$ and $\{X|X_j \geqslant s\}$, based on a threshold value $s$.

These steps are repeated on the new subspaces until a stopping criterion (e.g., minimum set size) is reached. For supervised classification, the optimal parameters for this tree (e.g., the stopping criterion, the structure of each tree) are chosen to minimize some predefined error function based on a set of preclassified training data. Individual decision trees can be highly variant, since the final classifications at the end of the tree depends on where splits are made. Random forests seek to reduce this variance by using statistical methods such as bootstrap aggregation (Breiman 1996).

The final classifications are determined by a majority vote of all of the decision trees. Random forests are a popular machine-learning algorithm as they are fast, can be parallelized, and do not require large training data sets to be constructed. They have been used to reliably predict M dwarf subtypes based on photometric colors (Hardegree-Ullman et al. 2019), perform star-galaxy photometric classification in wide-field and transient surveys (Miller et al. 2017; Clarke et al. 2020), and conduct other photometric classification tasks (e.g., Richards et al. 2011; Dubath et al. 2011; Bloom et al. 2012; Brink et al. 2013). Random forests have also been used in a regression form to infer atmospheric parameters from brown dwarf and exoplanet spectra (Márquez-Neila et al. 2018; Oreshenko et al. 2020).

We used the `RandomForestClassifier` implementation of the random forest algorithm provided in the `scikitlearn` package (Pedregosa et al. 2011). Our preclassified training data (all with $J$ S/N > 3) come from two sources: WISPS and 3D-HST spectra for non-UCD sources and the UCD template sample (Section 3.1). For the latter, we selected the 20 highest S/N spectra for each spectral subtype, smoothed and resampled these spectra to match the WFC3 resolution and wavelength range, and added Gaussian noise to encompass the range of $J$-S/N values present in the WFC3 data set down to the $J$ S/N = 3 limit. The training data we organized into five classes: a contaminants class composed of WFC3 spectra of nonpoint sources and point sources visually confirmed to not be UCDs (138,214 spectra total), M7–M9 dwarfs (4,686 spectra), L dwarfs (23,140 spectra), T dwarfs (13,479 spectra), Y dwarfs (12,871 spectra), and late-M/L subdwarfs (14,977 spectra)

Choosing an appropriate set of features to classify a data set is a critical design choice of any machine-learning model. We used the five spectral indices and four S/N statistics defined in Section 3.2.2, the best-fit spectral type obtained from the spectral template fitting, and the $F$-test statistic comparing the template and line fits. In cases where features were not measurable due to small/zero denominators (common for noisy spectra), we assigned an extreme value of −99 to indicate missing data. The initial preclassification data set was shuffled and split into a training set (75%) and a testing set (25%) for constructing, optimizing, and evaluating the random forest model.

The parameters describing the random forest model (Table 5) were optimized using the randomized search cross-validation algorithm (Bergstra & Bengio 2012). We used the $K$-fold cross-validation method in which the data is randomly split into $K$ equally sized subsamples, the model is fit to $K-1$ of these subsamples, and the model is then validated on the last subsample. The cross-validation algorithm conducts a random search over distributions of the tree parameters, weighting those with the smallest classification error highest. We assumed an initially uniform distribution of parameters, and utilized the default $K = 5$ cross validation implemented in `RandomizedSearchCV`.

Classification error, and hence model convergence, was determined by evaluating four metrics[9] that compare the rates of true positives (TP), true negatives (TN), false positives (FP), and false negatives (FN):

$$\text{accuracy:} \quad \mathrm{Ac} = (\mathrm{TP} + \mathrm{TN})/(\mathrm{TN} + \mathrm{TP} + \mathrm{FN} + \mathrm{FN}), \quad (8)$$

$$\text{precision:} \quad \mathrm{Pr} = \mathrm{TP}/(\mathrm{TP} + \mathrm{FP}), \quad (9)$$

[9] https://scikitlearn.org/stable/modules/model_evaluation.html

$$\text{recall:} \quad \text{Re} = \text{TP}/(\text{TP} + \text{FN}), \tag{10}$$

$$F_1 \text{ score:} \quad F_1 = (\text{Pr} \times \text{Re})/(\text{Pr} + \text{Re}). \tag{11}$$

These metrics were historically developed for information retrieval tasks in linguistics (Chinchor 1992). Accuracy measures the model's ability to correctly classify all sources, both positive and negative classifications. Precision measures the model's ability to distinguish true positives from false positives. Recall measures the model's ability to correctly select members of a particular class, and the $F_1$ score combines the precision and recall measures. We computed these metrics for individual classes and averaged across classes. Our model was constructed in the Google Colab environment[10] (Carneiro et al. 2018). The final model was selected as the one with the highest $F_1$ score in order to minimize the false negative rate and maintain accuracy; its parameters are summarized in Table 5.

Figure 6 displays the *confusion matrix* of our random forest model, the matrix of true and false positives and negatives as determined by the test set. Our best model has a >97% accuracy score across all classes, but varies between classes (Table 6). The precision, recall, and $F_1$ scores are generally above 90% with the exception of M7–M9 class precision. In terms of performance, there is little confusion between UCD classes; the percentage of M7–M9, L dwarfs, and T dwarfs classified as another UCD class is below 1%. The primary loss in accuracy and precision stems from early M dwarfs (<M7) being classified as galaxies, likely due to their weak molecular features.

Figure 6 shows the distribution of feature importance among the features used for classification. Feature importance captures the sensitivity of classification to a particular feature by measuring the decrease in classification error due to splits on that feature. The most important feature is the classification spectral type from the standard template fits. However, this feature alone cannot distinguish between contaminants and UCD types. For example, low S/N contaminants with strong emission features (active galaxies) are often classified as T dwarfs. Features combining $H_2O$ and continuum flux ratios therefore have high importance, as these are the highest S/N features and enable segregation of UCDs and contaminants. The least important features tend to be those sampling the noisiest parts of the data; e.g., $CH_4$ and $H_2O$-2 indices.

In traditional random forest models, the final classification of a source is based on the highest probability class across all decision trees in the random forest. To further reduce contamination, particularly in the M7–M9 dwarf class, we examined an alternative classification strategy in which classified UCDs were required to have classification probabilities >80%; all other sources were classified as contaminants. This 80% cutoff is determined from distribution of the class probabilities for UCDs selected by the index method.

Table 6 shows that this additional classification criterion generally increases the precision of detecting UCDs, particularly for the M7–M9 class, while significantly reducing the recall. In effect, this criterion produces a trade-off between minimizing contamination and maximizing the recovery of true UCDs. Given the rarity of UCDs in the overall sample, we find this trade-off to be warranted, as it can be modeled out when evaluating the overall sample (Paper II).

Following the optimization of our model parameters, we built and trained a final random forest model using the entire preclassified training set (with the 80% classification threshold), and applied it to the 46,561 WFC3 spectra of point sources with $J$ S/N > 3. In total, the random forest model classified 143 spectra as late-M, L, T, or Y dwarfs, including two objects identified as subdwarfs. Visual inspection confirmed 89 of these spectra as bona fide UCDs. The random forest classifier did not select 72 with spectral types of M7–L8 previously selected by the index selection as they fall below the threshold classification probability of 80%. On the other hand, the random forest method identified one additional M7 UCDs that was missed by the index selection. While finding fewer UCDs overall compared to the visual index selection, this method (at 80% probability cutoff) provided a more than tenfold reduction in contamination.

### *3.5. Selection by a Deep Neural Network*

We explored a second machine-learning classification approach using artificial neural networks (NNs; Schmidhuber 2014; LeCun et al. 2015; Shrestha & Mahmood 2019). Although the random forest method provided significant reduction in contamination, feed-forward artificial NNs outperform random forests in large-scale empirical studies (Caruana & Niculescu-Mizil 2006). In the simplest design, a multilayer, feed-forward NN with $L$ hidden layers, each of dimensionality $N$, is a set of functions that transform an input vector $\vec{X}$ of dimensionality $M$ into an output vector $f_k(\vec{X})$ of dimensionality $K$. Following the notation of Hastie et al. (2009), the set of transformations from $\vec{X}$ to $f_k(\vec{X})$ are given by

$$\vec{Z} = \sigma_0(\boldsymbol{\alpha} \cdot \vec{X}), \tag{12}$$

$$\vec{A_1} = \sigma_1(\boldsymbol{\beta}_1 \cdot \vec{Z}), \tag{13}$$

$$\vec{A_2} = \sigma_2(\boldsymbol{\beta}_2 \cdot \vec{A_1}), \tag{14}$$

$$\vdots \tag{15}$$

$$f_k(\vec{X}) = g_k(\vec{A_L}). \tag{16}$$

Here, $\sigma_0$, $\sigma_1$, $\sigma_2 \ldots \sigma_L$, and $g_k$ are scale functions also known as activation functions; while the matrices $\boldsymbol{\alpha}$ of dimension $N \times M$ and $\boldsymbol{\beta}$ of dimension $K \times N$ are the NN weights that linearly combine the various nodes between the layers. Training an NN entails optimizing the NN weights in order to minimize a classification convergence criterion $R(\boldsymbol{\theta})$ for a network model parameter set $\boldsymbol{\theta} = \{\boldsymbol{\alpha}, \boldsymbol{\beta}_1, \boldsymbol{\beta}_2, \ldots \boldsymbol{\beta}_L, \vec{\Gamma}\}$, where $\vec{\Gamma}$ are design parameters for the network (Table 7). A convergence criterion commonly used is cross entropy (Hastie et al. 2009),

$$R(\boldsymbol{\theta}) = -\sum_{m=1}^{M}\sum_{k=1}^{K} Y_k \log f_k(X_m), \tag{17}$$

where $Y_k$ represents the expected label for input variable $\vec{X}$. Minimizing $R(\theta)$ can be achieved numerically using a stochastic gradient descent algorithm (Kiefer & Wolfowitz 1952).

While the mathematics behind NNs have been understood for decades, the recent emergence of large data sets in science and industry, and significant improvements in computational hardware, have contributed to their increased application in general and toward astrophysical problems in particular. Examples include the use of convolutional neural networks (CNNs) to model stellar properties based on light curves (Blancato et al. 2020), to predict galaxy spectra from their broadband images (Wu & Peek 2020), and to discover extrasolar planets (Shallue & Vanderburg 2018); and the use of generative adversarial networks (GANs) to extract features from noisy galaxy images (Schawinski et al. 2017). Baron (2019) provides a review of machine-learning applications, including NNs, in astronomy.

[10] https://colab.research.google.com/

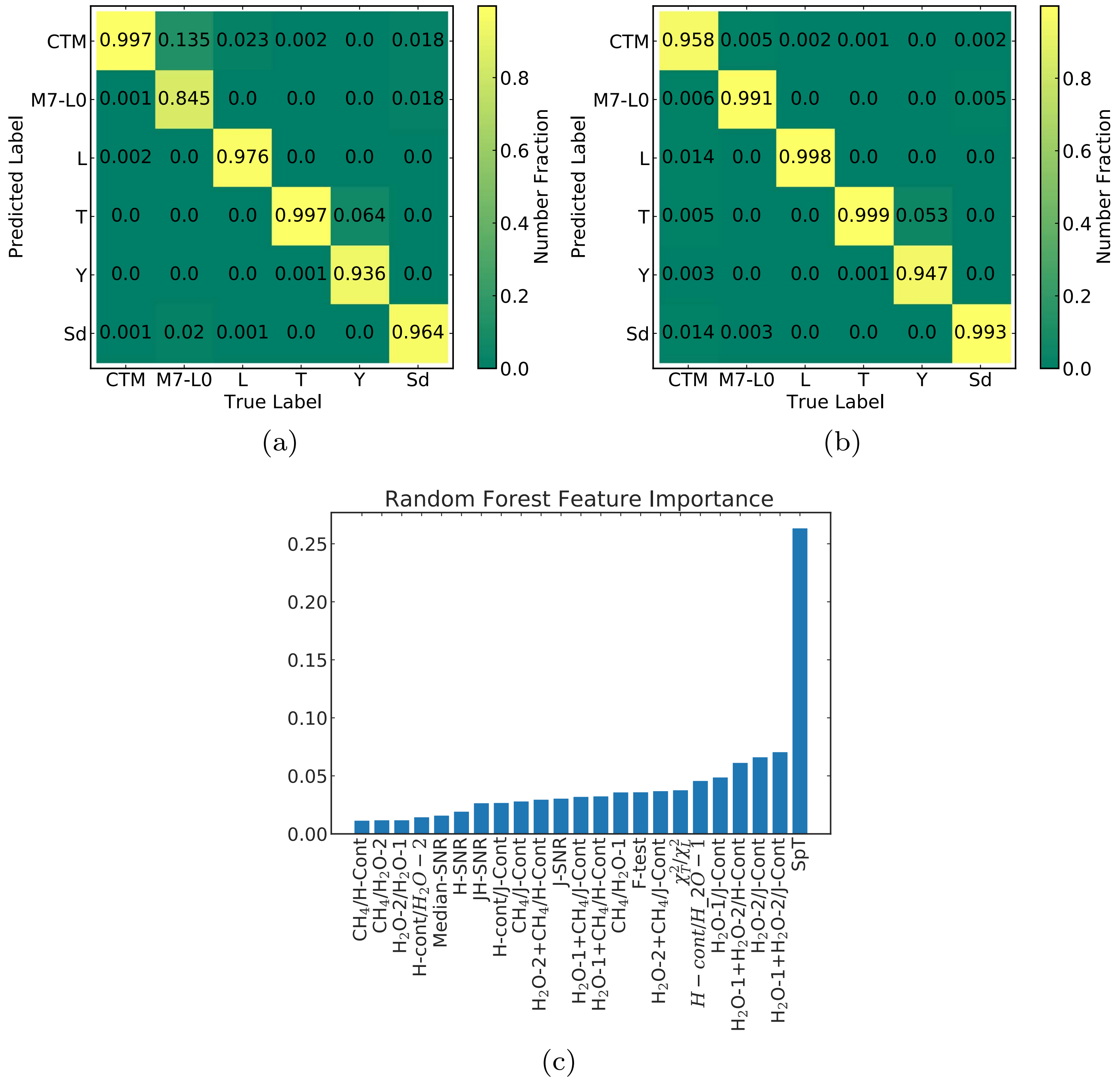

**Figure 6.** Top: random forest confusion matrix based on the test set. Numbers along the diagonal indicate the number of correctly classified spectra (true positives), while off-diagonals indicate false positives and false negatives. Predicted labels/classes in (a) were determined as the label with the maximum probability, while labels in (b) were determined by requiring a minimum classification probability for M7–M9, L dwarf, and T dwarf classes. The label Sd is used to denote the subdwarf class and the label CTM is used to denote contaminants. Bottom: bar plot showing the relative feature importance for classification on the training set. Low-importance features tend to sample the noisier regions of the spectra.

To train our NN, we used the same features and preclassified training set as the random forest classifier. Our NN model is a simple deep neural network (DNN) consisting of a series of fully connected layers (see Figure 7), implemented using the `Keras` library (Chollet et al. 2015). To converge to optimal parameters, we use adaptive moment estimation (Adam, Kingma & Ba 2014), a type of gradient descent algorithm. We used two forms of activation functions; for the input and hidden layers we used the rectified linear unit (ReLU, Nair & Hinton 2010) given by

$$\sigma(\vec{X})_i = \max(0, X_i), \tag{18}$$

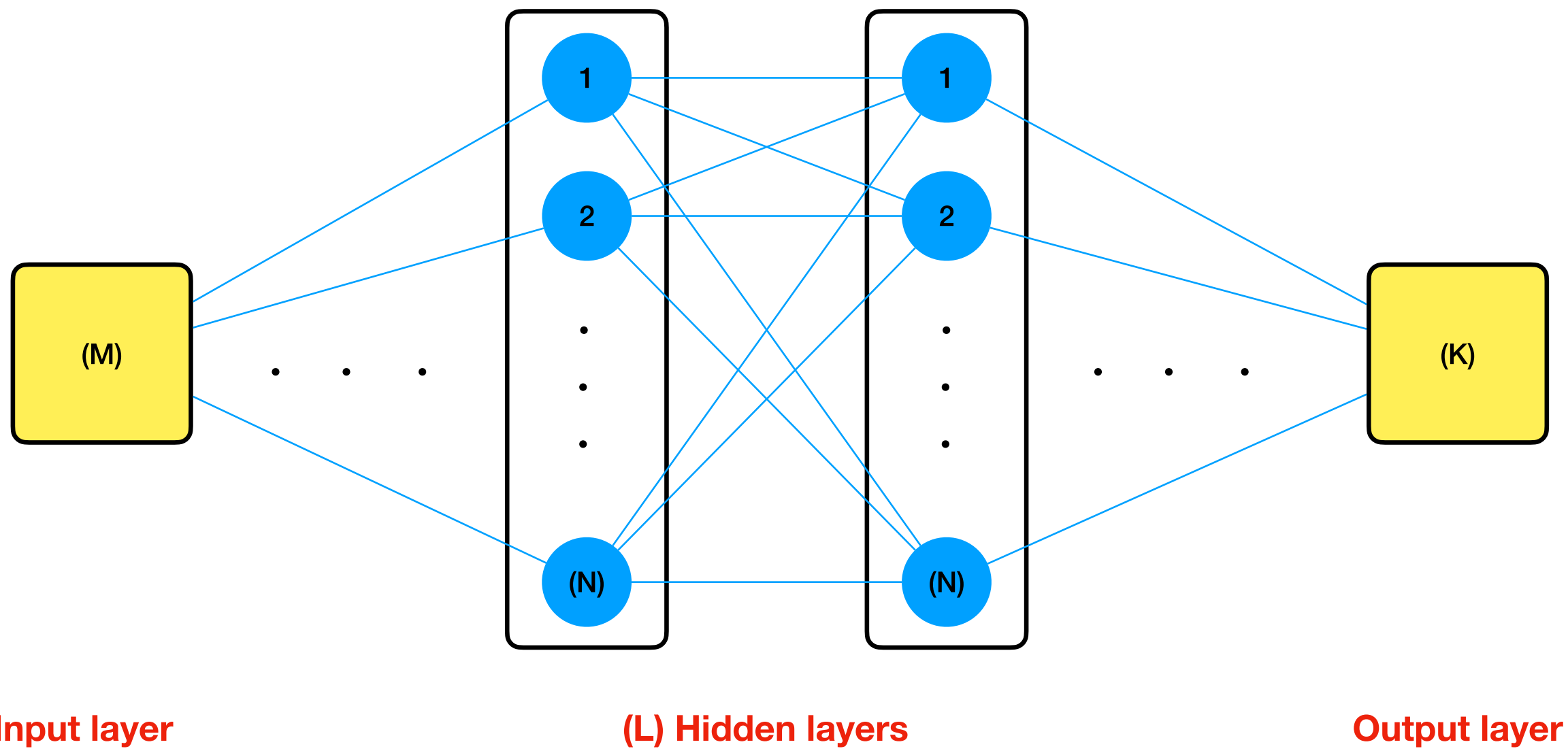

**Figure 7.** Schematic illustrating the DNN architecture used on this work. We summarize the model properties in Table 7

**Table 6**
Random Forest Metrics on Test Set

| Class | Accuracy | Precision | Recall | $F_1$ | TP | FP | TN | FN | Sample |
|---|---|---|---|---|---|---|---|---|---|
| Maximum probability[a] | | | | | | | | | |
| Contaminants | 0.991 | 0.997 | 0.989 | 0.993 | 34304 | 101 | 17055 | 382 | 34686 |
| M7–M9 | 0.994 | 0.845 | 0.918 | 0.880 | 1067 | 196 | 50484 | 95 | 1162 |
| L | 0.996 | 0.976 | 0.990 | 0.983 | 5666 | 141 | 45978 | 57 | 5723 |
| T | 0.996 | 0.997 | 0.933 | 0.964 | 3084 | 8 | 48528 | 222 | 3306 |
| Y | 0.996 | 0.936 | 0.999 | 0.966 | 3192 | 220 | 48428 | 2 | 3194 |
| Subdwarfs | 0.996 | 0.964 | 0.988 | 0.976 | 3724 | 139 | 47932 | 47 | 3771 |
| Classification probability >80%[b] | | | | | | | | | |
| Contaminants | 0.970 | 0.958 | 0.999 | 0.978 | 34662 | 1526 | 15630 | 24 | 34686 |
| M7–M9 | 0.995 | 0.991 | 0.799 | 0.885 | 929 | 8 | 50672 | 233 | 1162 |
| L | 0.990 | 0.998 | 0.910 | 0.952 | 5206 | 12 | 46107 | 517 | 5723 |
| T | 0.993 | 0.999 | 0.896 | 0.944 | 2961 | 4 | 48532 | 345 | 3306 |
| Y | 0.995 | 0.947 | 0.967 | 0.957 | 3090.000 | 172 | 48476 | 104 | 3194 |
| Subdwarfs | 0.990 | 0.993 | 0.862 | 0.923 | 3250.000 | 22 | 48049 | 521 | 3771 |

**Notes.**
[a] Predicted class is based on the highest probability class for that particular object.
[b] Predicted class must have a >80% classification probability; sources that do not satisfy this constraint are classified as contaminants.

**Table 7**
DNN Model Parameters

| Parameter | Description | Parameter Range | Best Value |
|---|---|---|---|
| L | Number of layers | (1, 10) | 10 |
| N | Number of nodes per layer | (1, 640) | 348 |
| $\eta$ | Learning rate | 0.1, 0.01, 0.001, and 0.0001 | 0.0001 |
| … | Batch size | … | 300 |
| … | Train, validation, test split | … | 0.4, 0.4, 0.2 |

while for the output layer we used the normalized exponential function (softmax) given by

$$\sigma(\vec{X})_i = \frac{e^{X_i}}{\sum_{j=1}^{M} e^{X_j}}. \tag{19}$$

The number of layers and nodes for the input and hidden layers are left as parameters to be determined through optimization.

To find the optimal parameters for the DNN model, we split our preclassified spectral data into a training set, a validation set, and a test set with proportions of 40%, 40%, and 20%. The training and validation sets are used to find and train the best model while the test set is used to evaluate the network performance on new data. We used a random search algorithm to optimize the model parameters listed in Table 7. We varied the number of nodes $N$ in each of the input and hidden layers, as well as the learning rate ($\eta$), which is a scale factor used to update model parameters from ($\theta$) to ($\theta'$).

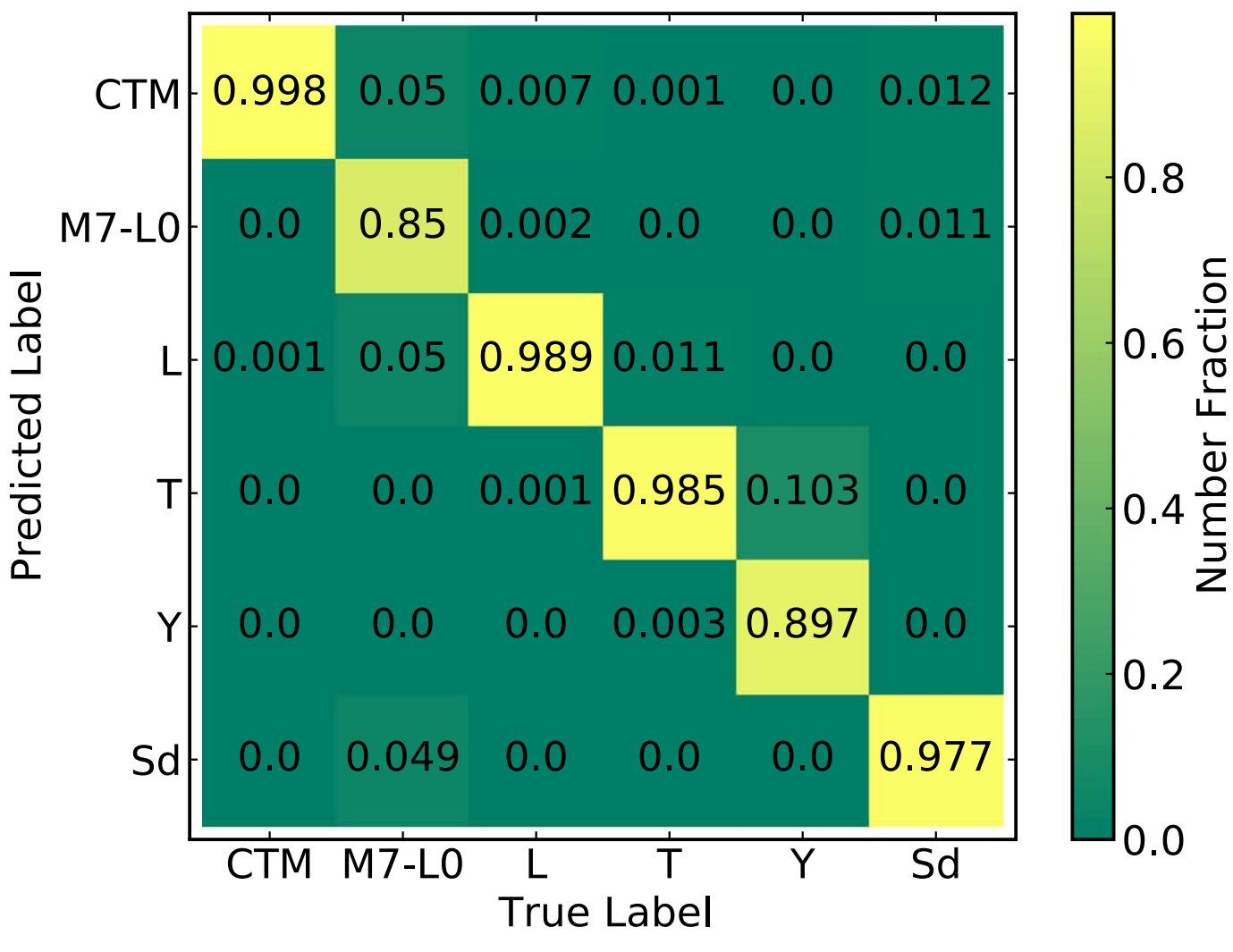

(a)

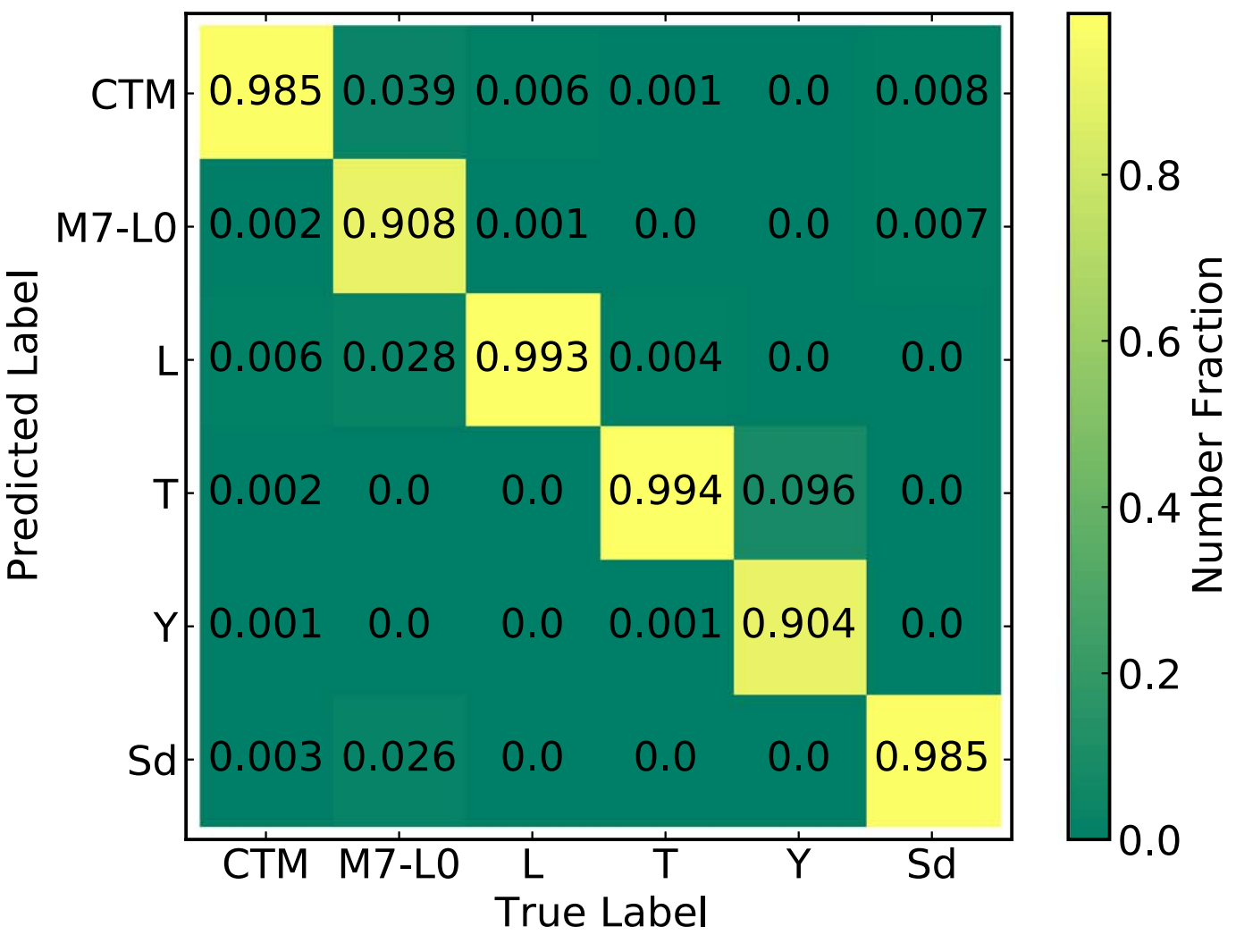

(b)

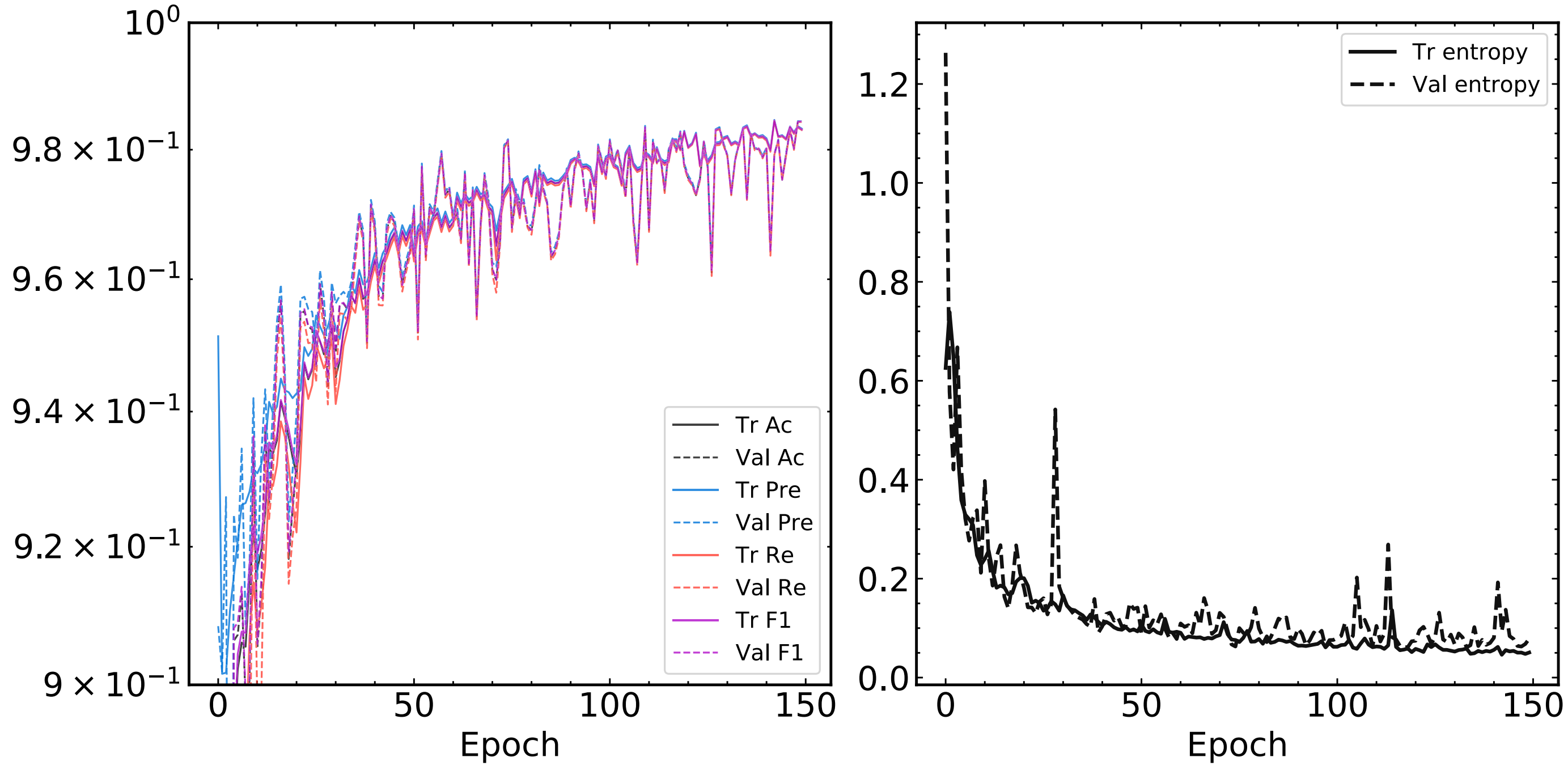

(c)

**Figure 8.** Top: confusion matrix for the DNN showing the performance on the test set. Predicted labels in (a) are computed from the maximum probability while labels (b) are computed by first making our selection cuts and then computing the maximum probability. Here, objects classified as <M7 and/or not satisfying our selection cuts are labeled as contaminants. The label Sd is used to denote the subdwarf class and the label CTM is used to denote contaminants. Bottom: DNN performance metrics of accuracy (Ac), precision (Pre), recall (Re), cross entropy, and $F_1$ on the training (Tr) and validation (Val) sets. We achieve high performance after roughly 120 epochs. Beyond 120 epochs, the performances of the validation and training sets start to deviate.

The parameter optimization was done on data samples (batches) of 300 spectra for each training epoch. We evaluated the same four performance metrics as used for the random forest (accuracy, precision, recall, and $F_1$ score), as well as cross entropy, which was used to optimize the performance of the DNN model. As shown in Figure 8, model convergence as traced by the performance metrics occurs within 150 update cycles. Our best-performing model has $L = 6$ hidden layers of $N = 64$ nodes each, for a total of 118,085 trainable weights. Analysis of the test set indicates an overall accuracy, precision, recall, and $F_1$ score of ~98% (Table 8).

After rebuilding the DNN model from the optimal parameters and training it with the full preclassified training set, we determined class probabilities for each of the six classes—contaminants, M7–M9 dwarfs, L dwarfs, T dwarfs, Y dwarfs, and subdwarfs—for the full WFC3 spectral sample. To reduce the number of false positives, we again imposed a classification probability threshold of >80% on the UCD

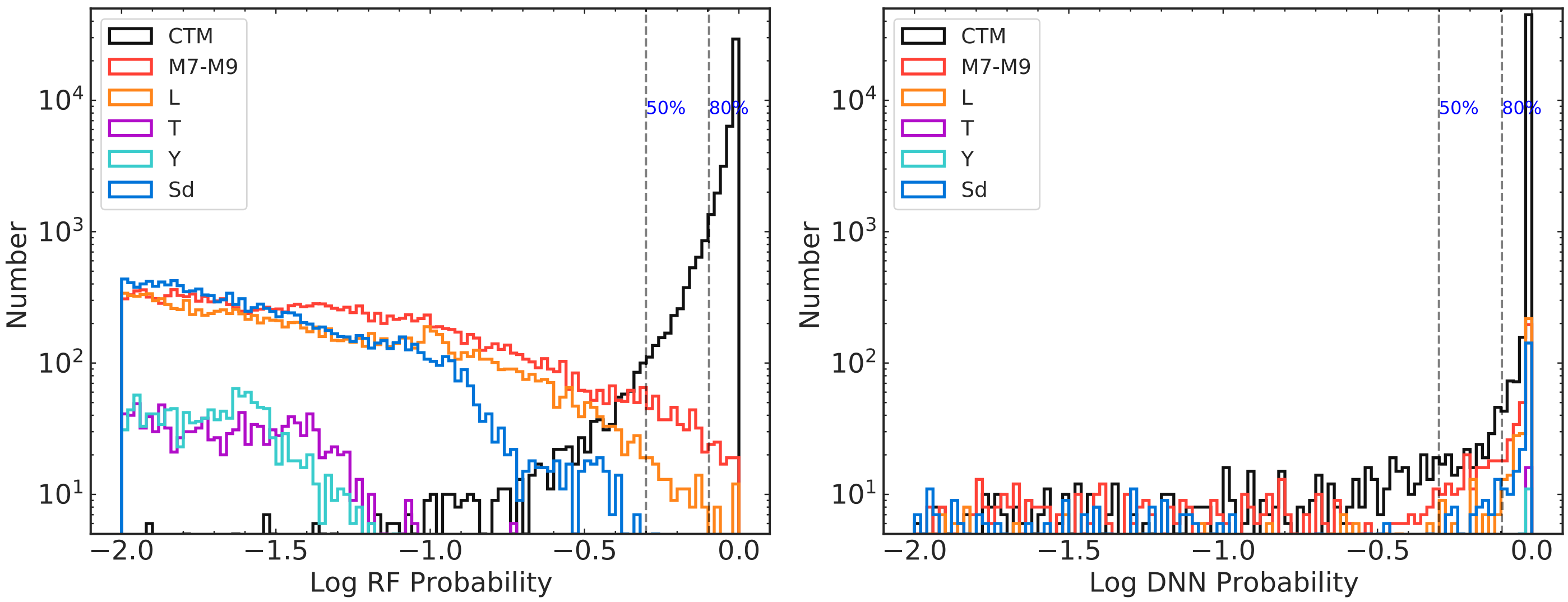

**Figure 9.** Log probability distributions of the random forest and NN models on the prediction set for each class. We mark the 50% and 80% cutoff as dashed lines. The majority of sources have low probabilities of being UCDs and high probabilities of being contaminants. The label Sd is used to denote the subdwarf class and the label CTM is used to denote contaminants.

**Table 8**
DDN Metrics on Test Set

| Class | Accuracy | Precision | Recall | $F_1$ | TP | FP | TN | FN | Sample |
|---|---|---|---|---|---|---|---|---|---|
| Maximum probability[a] | | | | | | | | | |
| Contaminants | 0.996 | 0.998 | 0.996 | 0.997 | 27,594 | 62 | 13,696 | 122 | 27716 |
| M7–M9 | 0.944 | 0.850 | 0.944 | 0.895 | 876 | 154 | 40,392 | 52 | 928 |
| L | 0.975 | 0.989 | 0.975 | 0.982 | 4532 | 49 | 36,778 | 115 | 4647 |
| T | 0.886 | 0.985 | 0.886 | 0.933 | 2359 | 36 | 38,776 | 303 | 2662 |
| Y | 0.997 | 0.897 | 0.997 | 0.944 | 2584 | 297 | 38,585 | 8 | 2592 |
| Subdwarfs | 0.978 | 0.977 | 0.978 | 0.977 | 2864 | 67 | 38,478 | 65 | 2929 |
| Classification probability >80%[b] | | | | | | | | | |
| Contaminants | 0.997 | 0.985 | 0.997 | 0.991 | 27,630 | 411 | 13,347 | 86 | 27716 |
| M7–M9 | 0.913 | 0.908 | 0.913 | 0.910 | 847 | 86 | 40,460 | 81 | 928 |
| L | 0.954 | 0.993 | 0.954 | 0.973 | 4431 | 33 | 36,794 | 216 | 4647 |
| T | 0.875 | 0.994 | 0.875 | 0.931 | 2329 | 14 | 38,798 | 333 | 2662 |
| Y | 0.990 | 0.904 | 0.990 | 0.945 | 2565 | 273 | 38,609 | 27 | 2592 |
| Subdwarfs | 0.960 | 0.985 | 0.960 | 0.973 | 2813 | 42 | 38,503 | 116 | 2929 |

**Notes.**
[a] Predicted class is based on the highest probability class for that particular object.
[b] Predicted class must have a >80% classification probability; sources that do not satisfy this constraint are classified as contaminants.

classes. Figure 9 displays the probability distribution of point sources in the full WFC3 spectral sample. We identified a total of 537 UCD candidates, including 158 objects classified as subdwarfs, of which 128 were visually confirmed. Like the random forest classifier, the DNN classifier provided a substantial reduction in contamination, but missed 36 UCDs identified by the index selection. On the other hand, the DNN classifier identified three additional UCDs (with spectral types of M7-M8) that were missed by the other two methods.

Combining the outcomes of all three selection methods, we identified 164 unique UCDs in the WISPS and 3D-HST surveys, listed in Tables 3 and 4 and subdivided into half-class subgroups in Table 9. An example of the field image, 2D spectral image, and 1D spectrum of one of our L dwarf discoveries is shown in Figure 10; see the associated figure set for the complete set of HST/WFC3 data for our sample. While each method selected roughly the same number of UCDs, the index-selection method had over 10 times the number of contaminants compared to the random forest classifier and over five times the number of contaminants compared to the DNN classifier. The contamination rate was largely reduced by the application of a classification probability threshold, which cannot be equivalently computed for the index selection method. In total, we found 113 UCDs in the WISPS fields and 51 UCDs in 3D-HST. We also identified an additional 83 sources that were classified earlier than spectral type M7 based on comparison to spectral standards, but matched to high S/N UCD spectral templates classified M7 or later (Section 3.1). These sources, listed in the Appendix, are not included in the main sample but are listed here as they likely to be mid- to late-M dwarfs.

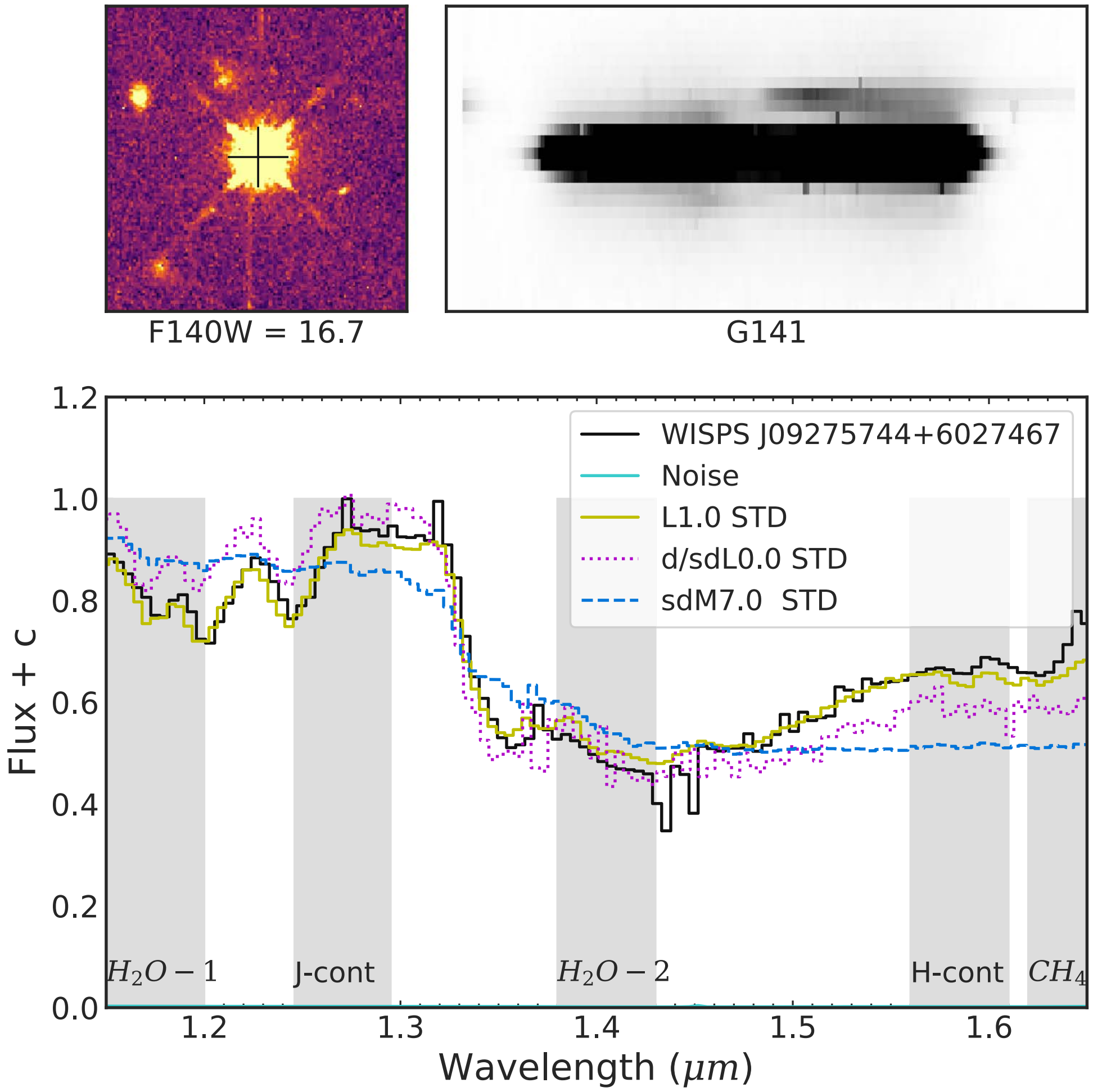

**Figure 10.** HST/WFC3 data for the UCD discovery WISPS J0925744+6027467. The top left panel displays a cutout of the F140W image of the field near the object. The top right panel shows a cutout of the G141 2D spectral image. The 1D extracted spectrum is shown on the bottom panel (black) compared to the best-fit dwarf (yellow solid line), best-fit mild subdwarf (d/sd; magenta dotted line), and best-fit subdwarf (sd; blue dashed line) standards. A figure set for all 164 sources in our sample is available in the online journal.

(The complete figure set (164 images) is available.)

**Table 9**
Number of UCDs Identified in the 3D-HST and WISPS Surveys

| Selection | Initial | Visually Confirmed | | | | | |
|---|---|---|---|---|---|---|---|
| | | M7–M9 | L0–L4 | L5–L9 | T0–T4 | T5–T9 | Total |
| Index selection | 3400 | 125 | 21 | 5 | 7 | 3 | 161 |
| Random forest (80% probability cut) | 143 | 62 | 13 | 4 | 7 | 3 | 89 |
| DDN (80% probability cut) | 537 | 101 | 15 | 4 | 7 | 3 | 128 |
| Random forest (50% probability cut) | 550 | 120 | 21 | 5 | 7 | 3 | 156 |
| DDN (50% probability cut) | 753 | 124 | 19 | 5 | 7 | 3 | 158 |
| Total (any method) | ⋯ | 128 | 21 | 5 | 7 | 3 | 164 |

## 4. Sample Characterization

### 4.1. Spectral Classifications

Classifications for our targets are based on comparisons to spectral standards using the method described in Section 3.2.3. We used the best-fit matches to determine the spectral types. Classification uncertainties were estimated using a weighted standard deviation, with weights computed from the $F$-test cumulative density distribution $W_i = F(\chi^2_{T,i}/\min(\{\chi^2_T\}),$ DOF, DOF), with $\chi^2_T$ defined in Equation (1). The median classification uncertainty is 1.4 subtypes, but varies between 0.5 and 9 subtypes. The largest classification uncertainties arise from the lowest S/N spectra, particularly for M-type UCDs. Figure 12 displays the distribution of spectral types, showing that our sample is heavily weighted toward late-M dwarfs (128 sources), as expected given their relatively bright intrinsic magnitudes and the corresponding larger volumes sampled in this magnitude-limited survey. We identified 26 L dwarfs (primarily L0–L4) and 10 T dwarfs, but no Y dwarfs or subdwarfs.

**Table 10**
Absolute Magnitude and Color versus Spectral Type Relations Used in this Work

| $x$ | $y$ | rms | Coefficients | | | | | | |
|---|---|---|---|---|---|---|---|---|---|
| | | | c6 | c5 | c4 | c3 | c2 | c1 | c0 |
| SpT[a] | WFC3 $M_{F110W}$ | 0.32 | −1.02260887E-06 | 1.78936774E-04 | −1.25922122E-02 | 4.57612828E-01 | −9.0880526E+005 | 9.41111020E+01 | −3.88061925E+02 |
| SpT | WFC3 $M_{F140W}$ | 0.37 | −1.26911098E-06 | 2.18686594E-04 | −1.51959378E-02 | 5.46285038E-01 | −1.07472467E+01 | 1.10362962E+02 | 4.52949443E+02 |
| SpT | WFC3 $M_{F160W}$ | 0.40 | −1.47948519E-06 | 2.52178047E-04 | −1.74232042e-02 | 6.25600790E-01 | −1.23318793E+01 | 1.26978487E+02 | −5.24118111E+02 |
| SpT | 2MASS J-F110W | 0.084 | −8.51745562E-08 | 1.56700496E-05 | −1.16134203E-03 | 4.43613741E-02 | −9.19918196E-01 | 9.78676767E+00 | −4.20410372E+01 |
| SpT | 2MASS J-F140W | 0.12 | −2.36588003E-07 | 3.91245500E-05 | −2.62006105E-03 | 9.08427011E-02 | −1.7201058E+00 | 1.68859410E+01 | −6.70968425E+01 |
| SpT | 2MASS H- F160W | 0.097 | −7.35527084e-08 | 1.20431676E-05 | −8.06167351E-04 | 2.82495698E-02 | −5.45254603E-01 | 5.45750570E+00 | −2.22166138E+01 |

**Note.** Empirical relations are computed as $y = \sum_{n=1}^{m} c_n x^n$ where $m$ is the power of the polynomial. Coefficients are written in decimal exponent notation.
[a] Decimal spectral type, with 10 = M0, 20 = L0, and 30 = T0, etc.

### *4.2. Color-spectral Type and Absolute Magnitude-spectral Type Relations*

All of the sources identified in the WISPS and 3D-HST surveys have apparent F110W, F140W, or F160W magnitudes from imaging data, which can be used in conjunction with source classifications to estimate distances. We determined new absolute magnitude/spectral type relations for M5–Y2 dwarfs in the F110W, F140W, and F160W filters by bootstrapping to the 2MASS *J*- and *H*-band relations of Pecaut & Mamajek (2013). For T7–Y2 dwarfs, we shifted to the absolute *H*-band magnitude/spectral type relation of Kirkpatrick et al. (2021).

We first computed color offsets between 2MASS (Vega magnitudes) and HST/WFC3 photometry (AB magnitudes) by computing spectrophotometric colors from our UCD template sample (Section 3.1). To ensure fidelity, we selected only those M5–T9 dwarfs with precise ( $\leqslant 1$ subtype uncertainty) red optical (late-M and L dwarfs) or infrared (late-L and T dwarfs) classifications; small 2MASS *J*- and *H*-band magnitude uncertainties ($\sigma < 0.4$); and spectra with median S/N > 100 for M dwarfs, S/N > 70 for L dwarfs, and S/N > 10 for T dwarfs. We also included all of the Y dwarfs from the template sample, and removed known binaries.

Magnitudes were computed by convolving the spectra with the appropriate filter profiles (HST[11]) or response functions (2MASS; Cohen et al. 1992), and uncertainties in spectral fluxes were propagated by random sampling. Figure 13 shows the resulting color offsets as a function of spectral type. Variations in color offsets are largest when the 2MASS and HST filters are more widely separated, reflecting the substantial evolution in spectral morphology across UCD spectral classes. We chose those colors that minimized spectral type variation, and hence bootstrapped the F110W and F140W relations to the *J* band and the F160W relation to the *H* band. Adding the relevant correction for each individual source to its absolute 2MASS magnitude from the absolute magnitude relation, we fit the entire spectral sequence with a sixth order polynomial using least-squares fitting, masking sources that were more than 5$\sigma$ outliers. Fit coefficients and corresponding absolute magnitude uncertainties are listed in Table 10. The scatter in these absolute magnitude relations is approximately 0.4–0.6 mag, comparable to other absolute magnitude/spectral type relations (e.g., Dupuy & Liu 2012; Best et al. 2018).

The 2MASS colors of discovered UCDs compared to colors of UCDs in our calibration samples are shown in Figure 11. These colors are computed using our relations in Table 10 and incorporating the intrinsic scatter in these relations. Our discoveries generally follow the spectral type/color trend from our calibration samples. However, objects with fairly large uncertainties in subtyping are significantly offset.

### *4.3. Distances and Galactic Coordinates*

Distance estimates were computed for each source based on the relations in Table 10 and measured HST/WFC3 photometry. We assume for simplicity that all sources are single, although magnitude-limited surveys of UCDs typically have a ∼20% multiplicity rate (Bouy et al. 2003; Burgasser et al. 2003, 2006b; Fontanive et al. 2018). Uncertainties in spectral classifications, source photometry, and the absolute magnitude relations were propagated by random sampling. In cases where more than one distance estimate was available from the absolute magnitude relations, we computed an uncertainty-weighted average. The distance of our UCD discoveries range over 60–3000 pc for late-M dwarfs, 40–2000 pc for L dwarfs, and 82–800 pc for T dwarfs (Figure 13). The most distant UCD with <50% distance uncertainty is the M7 ± 1 J23450092-4239288 at a distance of 2.1 ± 0.8 kpc. The closest sources are the L1.0 ± 0.7 WISPS J0927+6027 with an estimated distance of 48 ± 11 pc, and the M8 ± 1 WISPS J2333+3925 with an estimated distance of 67 ± 17 pc. Both of these sources are detected in Gaia with parallaxes of 19.497 ± 0.567 mas (51.4 ± 1.5 pc) and 19.498 ± 0.567 mas (50.5 ± 0.5 pc), respectively, such that WISPS J2333+3925 is in fact marginally closer. These are the only two sources in our UCD sample that are found in Gaia, and their measured and estimated distances are in agreement. Two of our extended sample targets (J10002943+0217108 and J14190738+5246567, Appendix) also have tentative Gaia detections, and their Gaia distances also agree with our photometric distance estimates within the uncertainties.

Figure 14 displays the distribution of source Galactic angular and spatial coordinates.[12] The angular distribution of the UCDs follows the HST pointings, with more sources identified in Galactic northern fields where there are more pointings. The average surface density of objects is 0.0813 arcmin$^{-2}$ at low Galactic latitudes ($15° \leqslant |b| \leqslant 30°$), 0.0801 arcmin$^{-2}$ at intermediate Galactic latitudes ($30° \leqslant |b| \leqslant 60°$), and 0.0625 arcmin$^{-2}$ at high Galactic latitudes ($|b| \geqslant 60°$). We observe 38 objects with $|Z| > 1$ kpc compared to 126 objects with $|Z| \leqslant 1$ kpc.

[11] https://www.stsci.edu/hst/instrumentation/wfc3/performance/throughputs

[12] We assume a right-handed, heliocentric Cartesian galactic coordinate system ($X$, $Y$, $Z$), where $X$ points toward the Galactic center ($l$, $b$) = (0°, 0°) , $Y$ points in the direction of Galactic rotation ($l = 90°$), and $Z$ points toward the north Galactic pole ($b = 90°$). For cylindrical coordinates ($R$, $Z$), we assume $R$ to be galactocentric, with the Sun located at $R_{\odot} = 8300$ pc and $Z_{\odot} = 27$ pc (Gillessen et al. 2009; Chen et al. 2001).

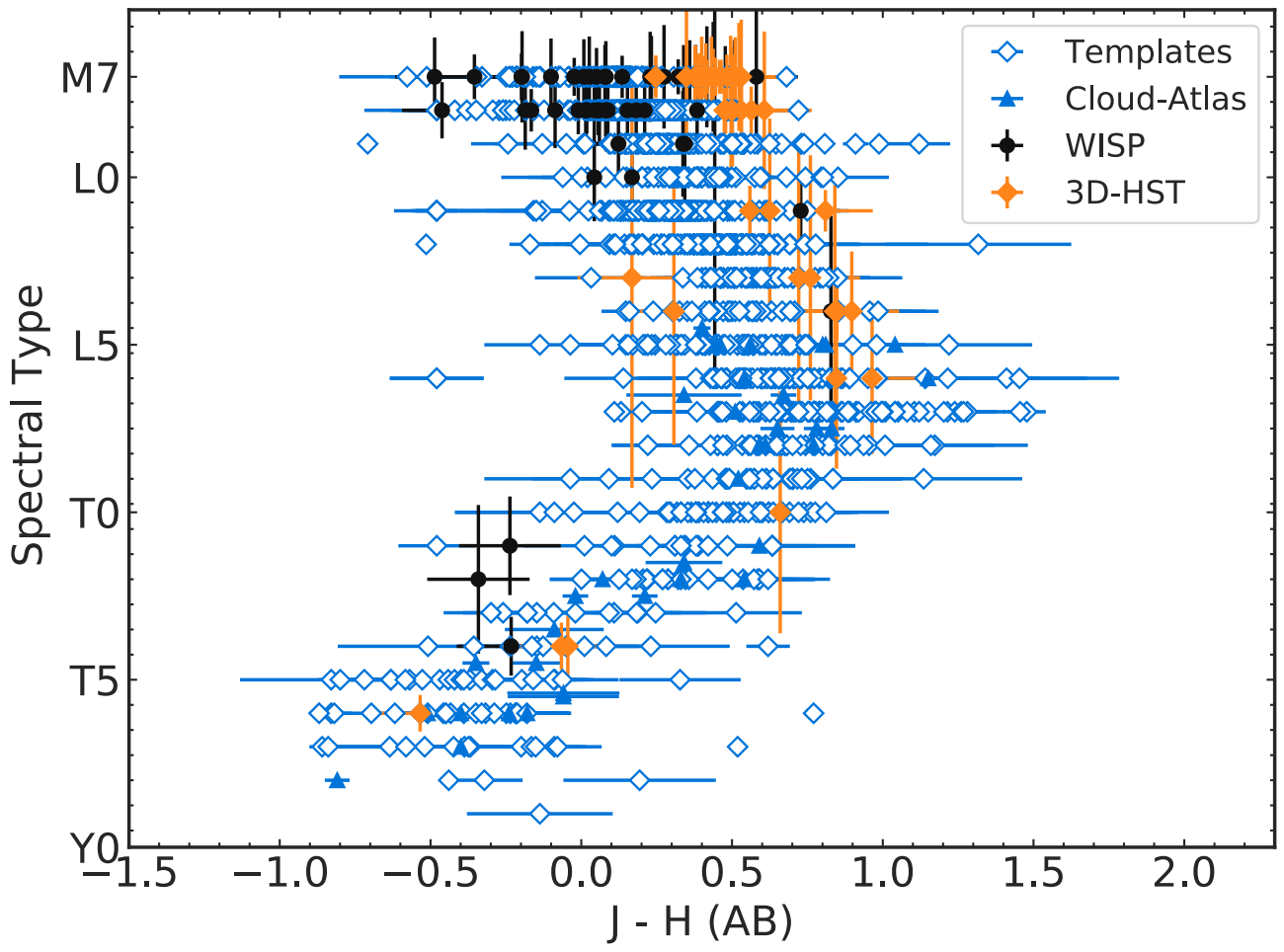

**Figure 11.** $J-H$ colors of UCDs in WISPS (black circles) and 3D-HST (orange diamonds) as a function of spectral type. These are compared to colors of sources in our template sample (blue diamonds) and from sources in the Cloud-Atlas sample (Manjavacas et al. 2017; blue triangles).

### 4.4. Detailed Discussion of Sample UCDs

*Late-M dwarfs*–We identified 128 sources with estimated spectral types of M7 $\leqslant$ SpT $<$ L0. Most of these classifications are robust, in that the major $H_2O$ absorption feature between 1.3–1.45 $\mu$m is clearly distinguishable from noise. For many of the WISPS spectra, we were able to confirm classification with G102 grism spectra (0.8–1.15 $\mu$m), which encompasses additional $H_2O$, FeH, and TiO absorption features characteristic of M dwarf spectra. As the G102 and G141 spectra sample distinct, non-overlapping wavelength ranges, we were unable to converge on a correct scaling to merge these spectra, so we retain the classifications based on the G141 spectra alone. We visually inspected all source images to confirm our UCDs are point sources unblended with other sources in the field, although a handful of sources with misaligned imaging data prevented this verification.[13]

One notable late-M dwarf discovery is the M8 ± 1.5 dwarf WISPS J2333+3921, which appears to have stronger $H_2O$ bands compared to its best-fit standard, which may be a signature of subsolar metallicity (Aganze et al. 2016). The G141 spectrum of WISPS J2333+3921 is better fit to a d/sd M9 standard despite its weaker FeH in the G102 spectrum. We also note that M7 ± 2 dwarf WISPS J1242+3538, located at an estimated distance of just over 3 kpc, is very close to an adjacent bright source and its spectrum may be partly contaminated. The closest M dwarf in our sample is the M8 ± 1 WISPS J2333+3925, as noted above at a Gaia parallax-based distance of 50.5 ± 0.5 pc. This source has a relatively high proper motion, $(\mu_\alpha, \mu_\delta)$ = (389.688 ± 0.203, −60.328 ± 0.178) mas yr$^{-1}$, corresponding to a tangential velocity of 94 km s$^{-1}$. It may be an old thin disk or a thick disk star.

*L Dwarfs*—We identified 26 L dwarfs, nearly half of which (12 sources) have classifications between L0 and L2; the distribution of sources per subtype throughout L0–L4 is nearly constant. Note that given the classification uncertainties, some of the earliest-type L dwarfs may be late M dwarfs, and similarly several classified late M dwarfs could be early L dwarfs. Most of these spectra are well matched to the L dwarf spectral standards, and several are confirmed by additional G102 grism data. As noted, the relatively nearby L2 ± 1.1 dwarf, WISPS J0927+6027, at a Gaia parallax-based distance of 51.4 ± 1.5 pc, is sufficiently bright (F140W = 16.697 ± 0.001) to make it amenable for further ground-based follow-up. The Gaia proper motion $(\mu_\alpha, \mu_\delta)$ = (−6.780 ± 0.550, 28.435 ± 0.544) mas yr$^{-1}$ corresponds to a modest tangential velocity of 7 km s$^{-1}$.

*T dwarfs*—T dwarfs are readily distinguished from other sources by their strong $H_2O$ and $CH_4$ absorption features. We identified 10 T dwarfs which match well with spectral templates. These include the three T dwarfs previously identified in WISPS by Masters et al. (2012): WISPS J0307-7243 (T4 ± 0.5 at ∼320 pc), WISP J1232-0033 (T7 ± 0.9 at ∼150 pc), and WISPS J1305-2538 (T8 ± 0.7 at ∼80 pc); and AEGIS J1418+5242 (T4 at ∼330 pc), first identified in 3D-HST data by Brammer et al. (2012). One of the early T dwarf discoveries, WISP J0326-1643 (T1 ± 1.4 at 740 pc), is also closely aligned with a brighter source, although its spectrum is a good fit to standards.

### 4.5. Ultracool Subdwarfs and Y Dwarfs

Despite creating specific selection criteria, we did not identify any obvious Y-type dwarfs or metal-poor subdwarfs in our sample. The lack of Y dwarfs is likely a consequence of the relatively small volume sampled for these intrinsically faint brown dwarfs. In Paper II, we estimate an expected number of Y dwarfs in this sample to be <1. The lack of subdwarfs may be due to their low abundance and selection biases. Ultracool subdwarfs are known to exist in the solar neighborhood (Burgasser et al. 2003; Zhang et al. 2019; Schneider et al. 2020), but are rare compared to solar-metallicity disk dwarfs (∼0.25% in the solar neighborhood; Pirzkal et al. 2009). However, subdwarf members of the halo and thick disk populations should be more common at the larger distances probed by this survey (see, Ryan & Reid 2016). Lodieu et al. (2017) measured a surface density of ⩾M5 subdwarfs of 0.04 deg$^{-2}$ to a depth of $J = 18.8$. For an average limiting depth of F110W = 22 this corresponds to ≈ 16 deg$^{-2}$, or roughly 10 ⩾ M5 subdwarfs over our total search area of ≈0.6 deg$^2$. Nevertheless, no subdwarfs were identified by our index selection, random forest classifier, or DNN classifier. There are several possible reasons that these sources were missing. First, subsolar metallicity is known to suppress the strength of molecular features, notably $H_2O$, in the near-infrared spectra of UCDs, largely through the presence of enhanced collision-induced $H_2$ absorption (Linsky 1969; Burgasser et al. 2007; Zhang et al. 2017). Relatively featureless late-M subdwarf spectra would have been rejected by our line fitting test, while more structured L subdwarf spectra would have been exceedingly rare in our sample. In Paper II, we model the full selection function of our search criteria applied to our calibration samples. Thus, a measurement of the density and distribution of ultracool subdwarfs likely requires the inclusion of the shorter-wavelength red optical data to capture characteristic features in this region (e.g., red optical slope and TiO, VO, FeH, and CrH bands), but the data remains incomplete for these short wavelengths.

[13] These sources were GOODSS J0332-2749, COSMOS J1000+0227, GOODSN J1237+6210, WISPS J2005-4139, WISPS J2038-2021, and WISPS J2248-8011.

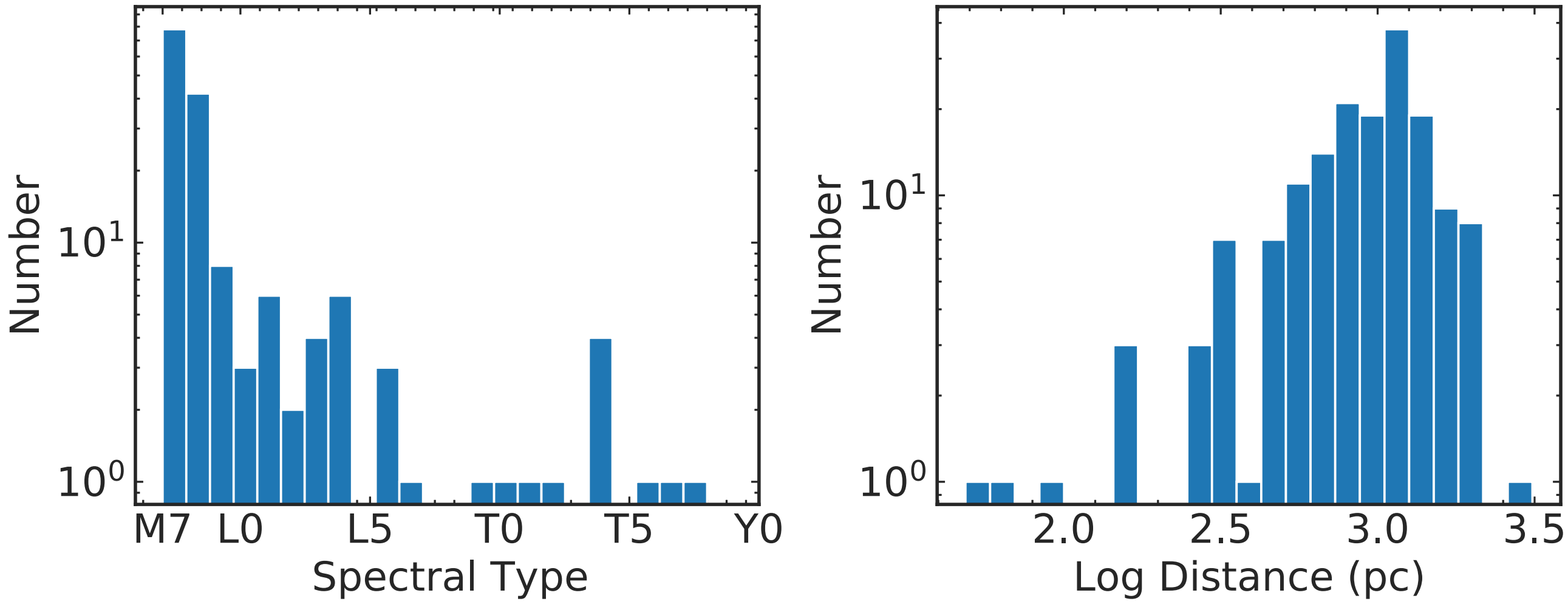

**Figure 12.** Spectral type and distance distributions of UCDs in the WISPS and 3D-HST surveys. Left panel: distribution of spectral types based on the spectral template fits. Right panel: distribution of spectrophotometric distances. The sample is dominated by late M dwarfs at distances $\gtrsim$1 kpc.

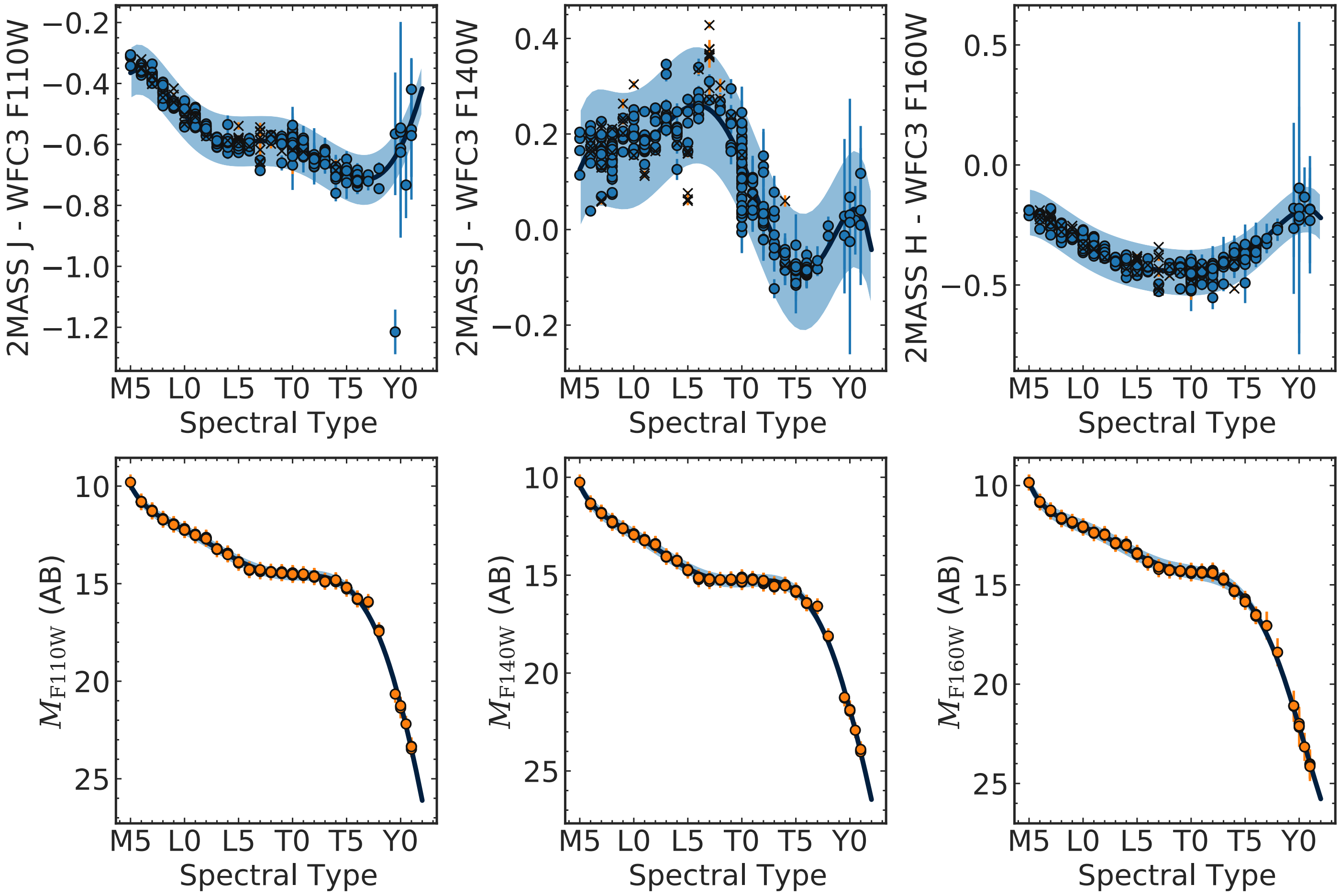

**Figure 13.** Top: offsets between 2MASS $J$- and $H$-band magnitudes and HST/WFC3 F110W, F140W, and F160W magnitudes as a function of spectral type, based on spectrophotometry of spectral templates. Bottom: derived absolute magnitude-spectral type relations for HST filters. Orange points show relations bootstrapped to 2MASS $J$-band photometry, blue points show relations bootstrapped to 2MASS $H$-band photometry. The solid lines delineate the sixth order polynomials whose coefficients are listed in Table 10.

## 5. Summary

We summarize our findings as follows:

1. We have identified 164 late-M, L, and T dwarfs in 0.6 deg$^2$ of HST/WFC3 parallels grism spectra from the

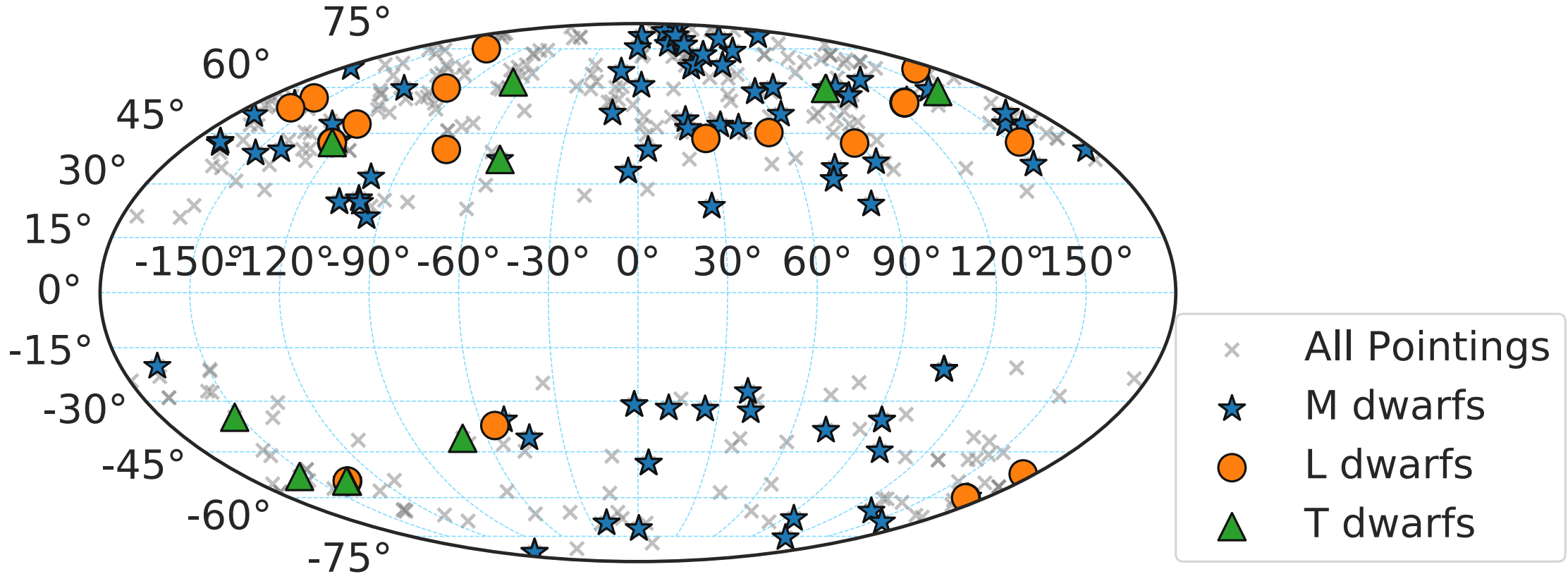

(ii)

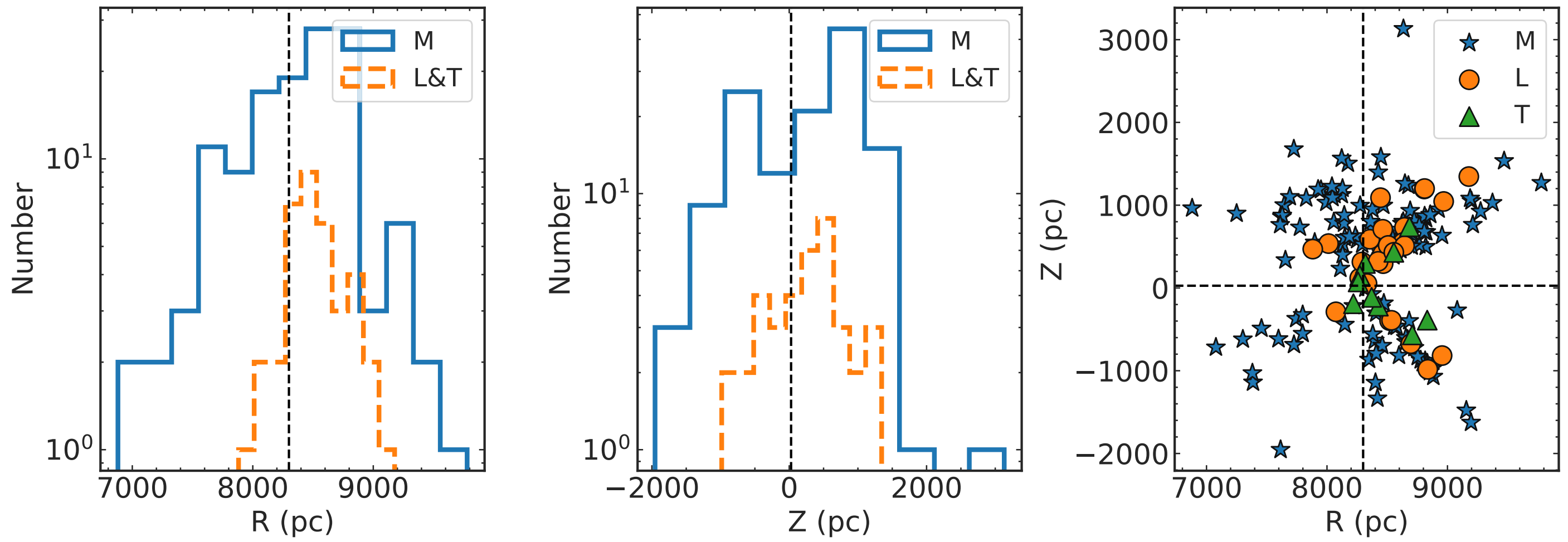

(i)

**Figure 14.** Top panel: distribution of observed UCDs in galactic angular coordinates ($l$, $b$). Gray X symbols indicate HST pointings without any UCDs. Bottom panel: distribution of observed UCDs in cylindrical galactic coordinates ($R$, $Z$; left and middle) and scatter plot of (right) these coordinates. Sources are color coded according to spectral type. Later-type UCDs are concentrated around the solar neighborhood while earlier-type M dwarfs are located further into the Galactic disk.

WISPS and 3D-HST samples. These objects were selected from over 250,000 sources using a combination of imaging morphology, spectral indices, template fitting, and visual confirmation.

2. We evaluated three different methods for source identification: traditional spectral index-based selection, random forest classification, and DDN classification. The latter two provided a substantial reduction in source contamination while missing only a small number of late-M dwarfs, and provide a superior approach to the discovery of rare sources in large spectral samples.
3. Spectral classification and empirical relations allow us to map the distances of these sources to up to 3 kpc, including many of the most distant spectroscopically confirmed brown dwarfs currently known.
4. The lack of detection of clear subdwarfs and/or Y dwarfs in the sample is expected due to the rarity of these populations and small area coverage of these surveys, although the limited wavelength range and our selection criteria may have biased the sample against late-M subdwarfs.

This paper provides an expansion of previous deep spectroscopic surveys, limited to only a handful of sources with $d \lesssim 1$ kpc. Nevertheless, the number of known distant L and T dwarfs remains small. Fortunately, forthcoming wide-field spectral surveys achieved with the James Webb Space Telescope (JWST; Gardner et al. 2006), SPHEREx (Doré et al. 2014), and Nancy Grace Roman Space Telescope (Spergel et al. 2015), and deep multi-epoch, multi-color photometry and astrometry with the Vera Rubin Observatory (Ivezić et al. 2019) and the Euclid space telescope (Laureijs et al. 2011) will allow us to probe to be greater depths ($J \approx 27$) and broader fields of view. Ryan & Reid (2016) have predicted a mean UCD surface density of $\approx 0.3$ arcmin$^{-2}$ (or $\approx$1000 deg$^{-2}$) in JWST down to a depth of $J = 24$. The recently approved JWST Parallel Application of Slitless Spectroscopy to Analyze Galaxy Evolution survey (PASSAGE, Cycle 1 GO-1571, PI Malkan) will obtained slitless grism spectra over a total area of 7.5 deg$^2$ down to F115W and F200W $\approx 27$ in a manner similar to WISPS and 3D-HST. We expect this survey to detect of order $10^4$–$10^5$ thin disk UCDs, thousands of thick disk UCDs, and hundreds of halo UCDs out to distances of 10 kpc, with full 1–2.2 $\mu$m spectra, enabling more robust segregation of contaminants and characterization of discoveries. In Paper II, we use this sample and a set of Monte Carlo simulations incorporating assumptions about the stellar mass function, star formation history, UCD evolutionary models, and Galactic structure to constrain the Galactic scale-height and population ages for these distant UCDs.

This research is based on observations made with the NASA/ESA Hubble Space Telescope obtained from the Space Telescope Science Institute, which is operated by the Association of Universities for Research in Astronomy, Inc., under NASA contract NAS 5-26555. These observations are associated with program(s) GO-12177 and GO-12328 (3D-HST), and programs GO-11696, GO-12283, GO-12568, GO-12902, GO-13352, GO-13517, and GO-14178 (WISPS). Support for this work was provided by NASA through the NASA Hubble Fellowship grant HST-HF2-51447.001-A awarded by the Space Telescope Science Institute, which is operated by the Association of Universities for Research in Astronomy, Inc., for NASA, under contract NAS5-26555. This research has benefited from the SpeX Prism Library, maintained by Adam Burgasser at http://www.browndwarfs.org/spexprism. C.A. thanks Lucianne Walkowicz, Adam Miller, Ivelina Momcheva, and members of the WISPS team for help in analyzing survey data; and the LSSTC Data Science Fellowship Program, which is funded by LSSTC, NSF Cybertraining grant No. 1829740, the Brinson Foundation, and the Moore Foundation. C.A. acknowledges funding from the UC Office of the President UC-HBCU Pathways Program. This material is based upon work supported by the National Aeronautics and Space Administration under grant No. NNX16AF47G issued through the Astrophysics Data Analysis Program. Portions of this work were conducted at the University of California, San Diego, which was built on the unceded territory of the Kumeyaay Nation, whose people continue to maintain their political sovereignty and cultural traditions as vital members of the San Diego community.

*Software*: Astropy (Price-Whelan et al. 2018), SPLAT (Burgasser & Splat Development Team 2017), SciPy (Virtanen et al. 2020), Matplotlib (Hunter 2007), Seaborn (Waskom et al. 2014), NumPy (Harris et al. 2020), Pandas (McKinney 2010), Keras (Chollet et al. 2015), Scikit-learn (Pedregosa et al. 2012), Axe (Kümmel et al. 2009).

## Appendix
## Additional Late-M Dwarf Discoveries

Here, we provide a table of 83 additional UCD candidates (Table 11) that were classified earlier than spectral type M7 based on comparison to spectral standards, but matched to high S/N UCD spectral templates classified M7 or later (Section 3.1). These sources were not included in the main sample, but are likely to be mid- to late-M dwarfs.

**Table 11**
Additional Late-type M dwarfs Identified in WISPS and 3D-HST Data

| Designation | Field ID | F110W | F140W | F160W | SNR-J | SpT | SpT Standard[a] | Distance (pc) | Selection[b] |
|---|---|---|---|---|---|---|---|---|---|
| J00053715-5012590 | PAR260-00025 | 21.454 ± 0.004 | ⋯ | 21.139 ± 0.007 | 32 | M7.0 ± 1.1 | M5.5 ± 0.7 | $1019^{+429}_{-262}$ | Indices |
| J01000226-5109007 | PAR198-00018 | ⋯ | 20.921 ± 0.009 | ⋯ | 14 | M7.0 ± 1.1 | M6.0 ± 1.9 | $658^{+318}_{-165}$ | Indices, DNN |
| J01063638+1508545 | PAR1-00003 | 18.312 ± 0.001 | 18.15 ± 0.001 | ⋯ | 162 | M7.0 ± 1.1 | M5.5 ± 0.7 | $221^{+103}_{-66}$ | Indices, DNN, RF |
| J01101162-0225002 | PAR84-00002 | 18.874 ± 0.001 | ⋯ | 18.919 ± 0.002 | 132 | M7.0 ± 1.1 | M5.5 ± 1.0 | $332^{+138}_{-81}$ | Indices, DNN, RF |
| J02171661-0510065 | UDS-19-G141-32382 | ⋯ | 21.916 ± 0.008 | 21.853 ± 0.006 | 16 | M7.0 ± 1.1 | M6.0 ± 1.7 | $1176^{+541}_{-321}$ | Indices, DNN |
| J04261278+0506145 | PAR251-00020 | ⋯ | 20.621 ± 0.009 | ⋯ | 20 | M7.0 ± 1.1 | M5.0 ± 1.7 | $580^{+263}_{-149}$ | Indices, DNN, RF |
| J04555058-2201253 | PAR194-00039 | ⋯ | 22.175 ± 0.031 | ⋯ | 7 | M7.0 ± 1.1 | M6.0: ± 2.8 | $1178^{+552}_{-318}$ | Indices, DNN, RF |
| J05011128-0548136 | PAR461-00034 | 21.326 ± 0.006 | ⋯ | 20.556 ± 0.007 | 19 | M7.0 ± 1.1 | M5.0 ± 1.4 | $875^{+420}_{-269}$ | Indices, RF |
| J05315500-0722222 | PAR321-00040 | 21.384 ± 0.003 | ⋯ | 21.59 ± 0.009 | 31 | M7.0 ± 1.1 | M5.0 ± 1.3 | $1118^{+441}_{-276}$ | Indices |
| J07183903+7425516 | PAR103-00017 | ⋯ | ⋯ | 20.586 ± 0.006 | 37 | M7.0 ± 1.1 | M6.0 ± 1.7 | $722^{+315}_{-172}$ | Indices |
| J08023317+1858595 | PAR188-00033 | ⋯ | 21.276 ± 0.01 | ⋯ | 14 | M7.0 ± 1.1 | M5.0: ± 2.3 | $774^{+358}_{-201}$ | Indices, DNN, RF |
| J09444166-1942144 | PAR293-00028 | ⋯ | 20.432 ± 0.007 | ⋯ | 43 | M7.0 ± 1.1 | M6.0 ± 2.0 | $524^{+259}_{-130}$ | Indices, DNN, RF |
| J10002943+0217108 | COSMOS-12-G141-10122 | ⋯ | 17.276 ± 0.0 | 17.16 ± 0.0 | 403 | M7.0 ± 0.5 | M6.0 ± 0.5 | $135^{+36}_{-30}$ | Indices, DNN, RF |
| J10030835+3245119 | PAR277-00011 | ⋯ | 20.119 ± 0.004 | ⋯ | 32 | M7.0 ± 1.1 | M6.0 ± 0.9 | $451^{+211}_{-118}$ | Indices, DNN, RF |
| J10053175-2304444 | PAR349-00059 | 22.07 ± 0.008 | ⋯ | 21.743 ± 0.013 | 14 | M7.0 ± 1.1 | M6.0 ± 1.3 | $1345^{+556}_{-326}$ | Indices, DNN |
| J10053651-2305456 | PAR349-00092 | 23.019 ± 0.013 | ⋯ | 22.896 ± 0.026 | 5 | M8.0 ± 1.1 | M6.0: ± 2.3 | $1850^{+604}_{-396}$ | Indices, DNN |
| J10053732-2304552 | PAR349-00025 | 20.588 ± 0.003 | ⋯ | 20.065 ± 0.004 | 41 | M7.0 ± 1.1 | M5.0 ± 1.3 | $653^{+295}_{-180}$ | Indices |
| J10054704+3705456 | PAR330-00048 | 21.712 ± 0.006 | ⋯ | 21.414 ± 0.011 | 20 | M7.0 ± 1.1 | M6.0 ± 1.8 | $1154^{+480}_{-279}$ | Indices, DNN |
| J10054713+3705460 | PAR330-00047 | 21.701 ± 0.006 | ⋯ | 21.574 ± 0.012 | 16 | M7.0 ± 1.1 | M6.0 ± 1.2 | $1190^{+462}_{-268}$ | Indices, DNN, RF |
| J10065615- 2952372 | PAR170-00053 | ⋯ | ⋯ | 21.181 ± 0.01 | 28 | M7.0 ± 1.1 | M5.0 ± 1.2 | $980^{+400}_{-245}$ | Indices |
| J10161887-2746264 | PAR375-00045 | 21.704 ± 0.007 | ⋯ | 21.393 ± 0.013 | 19 | M7.0 ± 1.1 | M6.5 ± 0.9 | $1162^{+452}_{-300}$ | Indices, DNN |
| J10244023+4811254 | PAR328-00028 | 21.233 ± 0.015 | ⋯ | 21.198 ± 0.031 | 25 | M7.0 ± 1.1 | M5.5 ± 0.7 | $969^{+399}_{-219}$ | Indices, RF |
| J10464955+1302398 | PAR116-00048 | ⋯ | ⋯ | 22.199 ± 0.013 | 11 | M7.0 ± 1.1 | M6.0: ± 2.2 | $1529^{+628}_{-374}$ | Indices, DNN |
| J11442277+0713567 | PAR434-00015 | 20.688 ± 0.003 | ⋯ | 20.842 ± 0.008 | 30 | M7.0 ± 1.1 | M6.0 ± 0.9 | $795^{+304}_{-201}$ | Indices |
| J11465352-1410588 | PAR481-00013 | 20.393 ± 0.003 | ⋯ | 20.054 ± 0.005 | 61 | M7.0 ± 1.1 | M5.0: ± 7.3 | $614^{+262}_{-151}$ | Indices, RF |
| J11471330-1404012 | PAR177-00007 | ⋯ | 19.645 ± 0.005 | ⋯ | 39 | M7.0 ± 1.1 | M5.5: ± 1.1 | $360^{+171}_{-92}$ | Indices, RF |
| J11545199+1941049 | PAR338-00035 | 22.238 ± 0.01 | ⋯ | 21.947 ± 0.017 | 13 | L1.0 ± 1.1 | M3.0: ± 3.4 | $859^{+217}_{-183}$ | Indices, DNN, RF |
| J12014072-1355480 | PAR121-00008 | 18.998 ± 0.001 | ⋯ | 18.666 ± 0.002 | 78 | M7.0 ± 1.1 | M6.0 ± 0.9 | $323^{+133}_{-80}$ | Indices, DNN, RF |
| J12301583+8236144 | PAR228-00020 | ⋯ | 20.731 ± 0.007 | ⋯ | 30 | M7.0 ± 1.1 | M6.0 ± 1.8 | $611^{+294}_{-155}$ | Indices, DNN, RF |
| J12361389+6209397 | GOODSN-21-G141-04680 | ⋯ | 22.469 ± 0.007 | 22.471 ± 0.01 | 11 | M7.0 ± 1.1 | M6.0 ± 1.8 | $1536^{+739}_{-429}$ | Indices |
| J13041367+3109486 | PAR362- 00030 | 21.981 ± 0.007 | ⋯ | 21.807 ± 0.012 | 11 | M7.0 ± 1.1 | M6.0 ± 1.5 | $1342^{+533}_{-322}$ | Indices, DNN, RF |
| J13275551+5248331 | PAR195-00023 | ⋯ | 21.05 ± 0.011 | ⋯ | 14 | M7.0 ± 1.1 | M6.0: ± 2.3 | $718^{+322}_{-190}$ | Indices, DNN, RF |
| J13351230+0801511 | PAR443-00009 | 20.864 ± 0.004 | ⋯ | 20.819 ± 0.008 | 27 | M7.0 ± 1.1 | M4.0: ± 4.6 | $822^{+331}_{-192}$ | Indices |
| J13541257+1803076 | PAR361-00005 | 18.89 ± 0.001 | ⋯ | 18.961 ± 0.002 | 136 | M7.0 ± 1.1 | M6.5 ± 0.7 | $339^{+145}_{-81}$ | Indices, DNN, RF |
| J13541539+1801592 | PAR361-00004 | 18.534 ± 0.001 | ⋯ | 18.958 ± 0.002 | 145 | M7.0 ± 1.1 | M6.5: ± 1.0 | $316^{+137}_{-81}$ | Indices, DNN, RF |
| J14094055+2621021 | PAR15-00041 | 22.477 ± 0.015 | 22.195 ± 0.018 | ⋯ | 12 | M7.0 ± 1.1 | M6.0 ± 1.1 | $1499^{+638}_{-488}$ | Indices, DNN |
| J14190738+5246567 | AEGIS-13-G141-21995 | ⋯ | 17.831 ± 0.0 | 17.828 ± 0.0 | 324 | M7.0 ± 1.1 | M5.0 ± 1.2 | $182^{+88}_{-52}$ | Indices, DNN |
| J14193008+0607203 | PAR345-00076 | 23.128 ± 0.02 | ⋯ | 22.965 ± 0.038 | 7 | M7.0 ± 1.1 | M6.0 ± 1.7 | $2286^{+942}_{-548}$ | Indices, DNN |
| J14195718+5254197 | AEGIS-10-G141-15910 | ⋯ | 20.54 ± 0.003 | 20.342 ± 0.006 | 54 | M7.0 ± 1.1 | M6.0 ± 0.5 | $611^{+290}_{-165}$ | Indices, DNN, RF |
| J14200918+5254075 | AEGIS-07-G141-06088 | ⋯ | 20.355 ± 0.001 | 20.327 ± 0.001 | 64 | M7.0 ± 1.1 | M6.0 ± 0.5 | $578^{+288}_{-160}$ | Indices |
| J14204934+2541298 | PAR301- 00038 | ⋯ | 21.865 ± 0.021 | ⋯ | 9 | M7.0 ± 1.1 | M6.0: ± 2.0 | $1020^{+461}_{-263}$ | Indices, DNN |
| J14270086+2352422 | PAR346-00021 | 20.968 ± 0.003 | ⋯ | 20.93 ± 0.006 | 29 | M7.0 ± 1.1 | M6.5 ± 1.0 | $864^{+362}_{-211}$ | Indices, DNN, RF |
| J14271124+2631530 | PAR218-00035 | ⋯ | 21.527 ± 0.015 | ⋯ | 10 | M7.0 ± 1.1 | M6.0: ± 2.1 | $886^{+419}_{-230}$ | Indices, DNN |

**Table 11**
(Continued)

| Designation | Field ID | F110W | F140W | F160W | SNR-J | SpT | SpT Standard[a] | Distance (pc) | Selection[b] |
|---|---|---|---|---|---|---|---|---|---|
| J14373187-0150224 | PAR64-00011 | 19.816 ± 0.002 | ⋯ | 19.604 ± 0.003 | 58 | M7.0 ± 1.1 | M6.0 ± 0.6 | $487^{+197}_{-122}$ | Indices,RF |
| J14373187-0150225 | PAR66-00011 | 19.886 ± 0.002 | ⋯ | 19.691 ± 0.003 | 63 | M7.0 ± 1.1 | M6.0 ± 1.0 | $504^{+214}_{-122}$ | Indices, DNN, RF |
| J15121258+0127330 | PAR235-00024 | ⋯ | 20.622 ± 0.006 | ⋯ | 23 | M7.0 ± 1.1 | M6.0 ± 1.4 | $581^{+271}_{-147}$ | Indices |
| J15345401+1252095 | PAR457-00024 | 21.465 ± 0.005 | ⋯ | 21.718 ± 0.012 | 16 | M7.0 ± 1.1 | M6.0 ± 1.4 | $1153^{+482}_{-279}$ | Indices, DNN |
| J15345628+1251173 | PAR457-00011 | 20.419 ± 0.003 | ⋯ | 20.648 ± 0.006 | 42 | M7.0 ± 1.1 | M6.5 ± 0.8 | $712^{+312}_{-179}$ | Indices |
| J15372024+5432247 | PAR174-00013 | ⋯ | ⋯ | 20.855 ± 0.008 | 36 | M7.0 ± 1.1 | M5.0 ± 2.0 | $822^{+355}_{-198}$ | Indices, DNN, RF |
| J15425795-1050564 | PAR449-00008 | 19.242 ± 0.002 | ⋯ | 18.717 ± 0.002 | 132 | M7.0 ± 1.1 | M6.0 ± 1.0 | $351^{+148}_{-92}$ | Indices, DNN, RF |
| J15430225-1050564 | PAR449-00015 | 19.78 ± 0.002 | ⋯ | 19.553 ± 0.004 | 85 | M7.0 ± 1.1 | M6.0 ± 1.2 | $484^{+196}_{-116}$ | Indices, DNN, RF |
| J15514396+2006007 | PAR445- 00037 | 21.789 ± 0.006 | ⋯ | 21.249 ± 0.011 | 21 | M8.0 ± 1.1 | M6.0: ± 2.3 | $963^{+312}_{-240}$ | Indices, DNN, RF |
| J15514891+2006007 | PAR445-00032 | 21.464 ± 0.004 | ⋯ | 20.897 ± 0.009 | 24 | M7.0 ± 1.1 | M5.0 ± 1.6 | $967^{+400}_{-262}$ | Indices |
| J15562889+2108114 | PAR308-00050 | 22.481 ± 0.009 | ⋯ | 22.453 ± 0.02 | 6 | M7.0 ± 1.1 | M6.0 ± 1.9 | $1753^{+770}_{-430}$ | Indices, DNN |
| J16045806+1446257 | PAR240-00076 | ⋯ | 22.808 ± 0.03 | ⋯ | 8 | M7.0 ± 1.1 | M6.5: ± 1.1 | $1559^{+764}_{-394}$ | Indices, DNN, RF |
| J16084912+6013510 | PAR82-00039 | ⋯ | ⋯ | 21.402 ± 0.01 | 19 | M7.0 ± 1.1 | M6.0 ± 1.6 | $1067^{+455}_{-247}$ | Indices, DNN, RF |
| J16150007-0940005 | PAR370-00036 | 20.972 ± 0.004 | ⋯ | 20.827 ± 0.008 | 27 | M7.0 ± 1.1 | M5.0 ± 1.4 | $834^{+352}_{-195}$ | Indices, RF |
| J16150186-0939558 | PAR370-00031 | 20.775 ± 0.003 | ⋯ | 20.142 ± 0.004 | 38 | M7.0 ± 1.1 | M6.0 ± 1.0 | $693^{+329}_{-198}$ | Indices, DNN, RF |
| J17004605+2922098 | PAR137-00018 | ⋯ | ⋯ | 20.401 ± 0.006 | 43 | M7.0 ± 1.1 | M5.0 ± 1.9 | $659^{+321}_{-160}$ | Indices |
| J17004733+2923168 | PAR137-00027 | ⋯ | ⋯ | 20.793 ± 0.007 | 31 | M7.0 ± 1.1 | M6.0 ± 1.4 | $805^{+316}_{-196}$ | Indices |
| J17014073+6407419 | PAR377-00162 | 23.571 ± 0.018 | ⋯ | 23.111 ± 0.026 | 3 | L2.0 ± 3.0 | M3.0: ± 2.8 | $1347^{+755}_{-542}$ | Indices, DNN, RF |
| J18322505+5344355 | PAR124-00053 | 21.882 ± 0.008 | ⋯ | 21.776 ± 0.015 | 18 | M7.0 ± 1.1 | M6.5 ± 1.0 | $1302^{+540}_{-322}$ | Indices, DNN, RF |
| J20054103- 4139180 | PAR371-00055 | 21.543 ± 0.006 | ⋯ | 20.927 ± 0.008 | 20 | M7.0 ± 1.1 | M6.5 ± 1.0 | $989^{+458}_{-264}$ | Indices, RF |
| J20054148-4139216 | PAR371-00027 | 20.261 ± 0.002 | ⋯ | 20.184 ± 0.005 | 40 | M7.0 ± 1.1 | M6.0 ± 0.8 | $615^{+261}_{-147}$ | Indices, RF |
| J20054544-4140012 | PAR371-00100 | 22.563 ± 0.009 | ⋯ | 22.464 ± 0.018 | 7 | M7.0 ± 1.1 | M6.0 ± 1.7 | $1770^{+745}_{-436}$ | Indices, DNN |
| J20054641-4140192 | PAR371-00078 | 22.184 ± 0.009 | ⋯ | 22.131 ± 0.018 | 11 | M8.0 ± 1.1 | M6.0: ± 2.4 | $1291^{+400}_{-263}$ | Indices, DNN, RF |
| J20054663-4140084 | PAR371-00034 | 20.803 ± 0.003 | ⋯ | 20.624 ± 0.005 | 28 | M7.0 ± 1.1 | M5.0 ± 1.8 | $780^{+317}_{-192}$ | Indices, RF |
| J20220562-3111528 | PAR332-00036 | 20.475 ± 0.002 | ⋯ | 20.449 ± 0.005 | 32 | M7.0 ± 1.1 | M4.0: ± 5.0 | $683^{+294}_{-157}$ | Indices |
| J20220626-3112072 | PAR332-00058 | 21.032 ± 0.004 | ⋯ | 20.758 ± 0.008 | 27 | M7.0 ± 1.1 | M5.0 ± 1.4 | $844^{+347}_{-210}$ | Indices, RF |
| J21373549-1438240 | PAR369-00034 | 21.283 ± 0.004 | ⋯ | 20.745 ± 0.007 | 30 | M7.0 ± 1.1 | M5.5 ± 0.9 | $902^{+388}_{-241}$ | Indices, DNN, RF |
| J21374036-1437588 | PAR369-00048 | 22.098 ± 0.007 | ⋯ | 21.999 ± 0.015 | 13 | M7.0 ± 1.1 | M6.0 ± 1.3 | $1449^{+576}_{-342}$ | Indices, DNN, RF |
| J21391523-3825012 | PAR309-00013 | 20.217 ± 0.002 | ⋯ | 20.17 ± 0.004 | 41 | M7.0 ± 1.1 | M5.5 ± 0.8 | $608^{+246}_{-145}$ | Indices |
| J22052802-0017073 | PAR94- 00043 | 22.169 ± 0.007 | ⋯ | 21.993 ± 0.015 | 13 | M7.0 ± 1.1 | M6.0 ± 2.0 | $1450^{+609}_{-353}$ | Indices, DNN |
| J22221887+0937258 | PAR50-00017 | ⋯ | 20.891 ± 0.007 | ⋯ | 22 | M7.0 ± 1.1 | M5.0: ± 2.0 | $641^{+317}_{-168}$ | Indices |
| J22533412-0644564 | PAR473-00018 | 21.4 ± 0.007 | ⋯ | 21.501 ± 0.016 | 11 | M7.0 ± 1.1 | M5.0 ± 1.7 | $1080^{+441}_{-269}$ | Indices |
| J23071868+2111362 | PAR166-00004 | ⋯ | ⋯ | 16.607 ± 0.001 | 326 | M7.0 ± 1.1 | M6.0 ± 1.2 | $120^{+52}_{-31}$ | Indices, DNN, RF |
| J23352023-3536432 | PAR359-00020 | 20.797 ± 0.003 | ⋯ | 20.722 ± 0.006 | 31 | M7.0 ± 1.1 | M6.0 ± 1.0 | $788^{+341}_{-189}$ | Indices |
| J23450387-4239216 | PAR356-00013 | 20.308 ± 0.002 | ⋯ | 19.861 ± 0.003 | 66 | M7.0 ± 1.1 | M6.0 ± 1.0 | $580^{+261}_{-151}$ | Indices |

**Notes.** Table 11 is available in machine-readable format.
[a] Classification by using spectral standards.
[b] Selection methods are index-index selection (IND), random forest selection (RF), and deep neural network (DNN).

(This table is available in machine-readable form.)

## ORCID iDs

Christian Aganze https://orcid.org/0000-0003-2094-9128
Adam J. Burgasser https://orcid.org/0000-0002-6523-9536
Mathew Malkan https://orcid.org/0000-0001-6919-1237
Christopher A. Theissen https://orcid.org/0000-0002-9807-5435
Roberto A. Tejada Arevalo https://orcid.org/0000-0001-6708-3427
Chih-Chun Hsu https://orcid.org/0000-0002-5370-7494
Daniella C. Bardalez Gagliuffi https://orcid.org/0000-0001-8170-7072
Russell E. Ryan, Jr. https://orcid.org/0000-0003-0894-1588
Benne Holwerda https://orcid.org/0000-0002-4884-6756